\documentclass[reprint,prx,aps, floatfix]{revtex4-1}
\usepackage{amsmath,amssymb,mathtools}
\usepackage{graphicx}
\usepackage{bm}
\usepackage{physics}
\usepackage{booktabs,bbm}
\usepackage{xcolor}
\usepackage{dcolumn}
\usepackage{bm}
\usepackage{amssymb}
\usepackage{braket}
\usepackage{color}
\usepackage{comment}
\usepackage{gensymb}
\usepackage{soul} 
\setstcolor{red} 
\usepackage{epstopdf}
\usepackage{hhline}
\usepackage{tabularx}
\usepackage{xspace}
\usepackage{layouts}
\usepackage{lineno} 
\usepackage{placeins}

\newcolumntype{C}[1]{>{\centering\arraybackslash}m{#1}}
\newcolumntype{N}{@{}m{0pt}@{}}

\usepackage[pdfencoding=auto,psdextra]{hyperref}
\hypersetup{
  colorlinks=true,
  linkcolor=blue,
  citecolor=blue,
  urlcolor=blue
}

\usepackage{xr-hyper}
\begin{document}

\title{Polychronous Wave Computing: Timing-Native Address Selection in Spiking Networks}

\author{Natalia G. Berloff}
\affiliation{Department of Applied Mathematics and Theoretical Physics, University of Cambridge, Cambridge CB3 0WA, United Kingdom}
\email{N.G.Berloff@damtp.cam.ac.uk}

\begin{abstract}
Spike timing offers a combinatorial address space, suggesting that timing-based spiking inference can be executed as lookup and routing rather than as dense multiply--accumulate.
Yet most neuromorphic and photonic systems still digitize events into timestamps, bins, or rates and then perform selection in clocked logic.
We introduce Polychronous Wave Computing (PWC), a timing-native address-selection primitive that maps relative spike latencies directly to a discrete output route in the wave domain.
Spike times are phase-encoded in a rotating frame and processed by a programmable multiport interferometer that evaluates K template correlations in parallel; a driven--dissipative winner-take-all stage then performs a physical argmax, emitting a one-hot output port.
We derive the operating envelope imposed by phase wrapping and mutual coherence, and collapse timing jitter, static phase mismatch, and dephasing into a single effective phase-noise budget whose induced winner--runner-up margin predicts boundary-first failures and provides an intensity-only calibration target.
Simulations show that nonlinear competition improves routing fidelity compared with noisy linear intensity readout, and that hardware-in-the-loop phase tuning rescues a temporal-order gate from $55.9\%$ to $97.2\%$ accuracy under strong static mismatch.
PWC provides a fast routing coprocessor for LUT-style spiking networks and sparse top-1 gates (e.g., mixture-of-experts routing) across polaritonic, photonic, and oscillator platforms.
\end{abstract}

\maketitle

\section{Introduction}

Spiking neural networks (SNNs) can encode information in when spikes arrive, not only in how many spikes occur.
If $N$ distinct input channels each emit one spike within a decision window, then the arrival order alone supports up to $N!$ distinguishable patterns (ignoring ties and finite timing resolution), giving timing codes a combinatorial address space.
This is the intuition behind the look-up-table SNN (LUT-SNN) perspective emphasized in the Spiking Manifesto: the forward pass can be viewed as selecting a discrete address from spike timing and then performing a look-up, rather than updating a dense real-valued activation vector \cite{Izhikevich2025SpikingManifesto}.
Viewed this way, the spike-time pattern serves as a discrete address. Inference then depends on which neuron (port) wins and hence which stored response is retrieved, rather  than producing a dense continuous activation via VMM and digitizing later.
Our work adopts this LUT-centric view and implements the timing $\to$ index step with physical waves.
In neuroscience, a closely related idea is polychronization: reproducible, time-locked, while not necessarily synchronous,  spatiotemporal firing patterns supported by heterogeneous axonal delays and plasticity \cite{Izhikevich2006Polychronization,Pauli2018ReproducingPolychronization}.

The difficulty is turning this combinatorial timing space into hardware throughput \cite{Wang2013FPGApolychronous}.
Many neuromorphic and photonic systems only compare timings after converting events into timestamps, clock bins, or accumulated rates, and then reducing them in clocked logic.
When the underlying device physics is ultrafast (picosecond--nanosecond), this digitize--compare--select pipeline can dominate latency and energy: the substrate evolves quickly, but the decision is serialized through conversion and reduction.
This mismatch is particularly relevant in exciton--polariton settings, where picosecond-scale delayed nonlinearities enable ultrafast temporal primitives \cite{Topfer2020TimeDelayPolaritonic,Mirek2022TimeDelayedNN} and polaritonic spiking dynamics operate on intrinsic $\mathcal{O}(10$--$100)$~ps lifetimes with sub-pJ-scale per-spike energy estimates \cite{Tyszka2023LIFPolaritons}.

We, therefore, move the comparison step into the physical substrate and digitize only once, after parallel scoring.
We call the resulting approach \emph{polychronous wave computing} (PWC), which acts as a hardware look-up module for spike timings. The only required encoding is a phase-referenced write of each event; no per-event timestamp is formed.
Within a decision window, spike times are mapped to phases in a rotating frame and fed to a programmable multiport interferometer that evaluates many timing patterns in parallel by coherent interference, producing complex template scores. 
A driven--dissipative mode-competition stage then performs winner-take-all (WTA) selection and emits a single one-hot output port, yielding one discrete address.
Operationally, PWC performs content-addressed lookup on spike-time patterns: it returns the address of the best-matching temporal hypothesis without per-spike timestamping or digital delay subtraction.

Two constraints set the basic operating envelope for phase-coded temporal lookup: the decision window must fit within a single unwrapped phase cycle, and coherence must persist across that window.
Within this envelope, timing uncertainty enters as an effective phase-noise budget.
A key practical consequence is that robustness is controlled by the separation between the best-matching output and its nearest competitor (a winner--runner-up margin): errors concentrate in near-tie events rather than appearing as a uniform loss of contrast.
Because margins are observable at the outputs, the same viewpoint suggests an intensity-only hardware-in-the-loop calibration loop that tunes programmable phases to enlarge margins and compensate static offsets and drift.

\noindent\textbf{Related work.}
Hardware approaches to temporal, spiking pattern processing broadly fall into three routes.
(i) \emph{Time-to-digital or  binned pipelines} convert events to timestamps, bins, or rates and then compute similarity and an argmax in clocked logic; they can implement discrete routing, but the comparison and reduction are not performed at the native physics timescale.
(ii) \emph{Photonic correlators or matched filters} compute similarity by interference \cite{VanderLugt1964,Goodman2015StatisticalOptics}, but they typically output an analog correlation peak or intensity distribution that must still be reduced to a stable one-hot address, and they do not by themselves provide calibrated winner--runner-up margins for robust routing.
(iii) \emph{Reservoir or delayed-nonlinearity time processing} exploits dissipative memory and rich dynamics for sequence processing \cite{Brunner2013,Mirek2021NanoLett,Tyszka2023LIFPolaritons}, but it does not provide a programmable, wrap-controlled temporal hypothesis scorer coupled to a physical argmax.

PWC targets the missing primitive: timing-native address selection.
Within a wrap-free, phase-coherent window, it evaluates $K$ temporal hypotheses in parallel by programmable interference and then digitizes the result into a one-hot output port via driven--dissipative competition.

More broadly, most optical and photonic ML hardware to date emphasizes fast linear algebra: coherent vector--matrix products and amplitude-encoded matrix--vector multiplication for real-valued activations \cite{Shen2017NatPhoton,Hamerly2019,Wetzstein2020Nature,Shastri2021,kalinin2023analog,kalinin2025analog}, or photonic spiking, neurosynaptic primitives \cite{Feldmann2019Nature,Bogaerts2020Nature}.
By contrast, our module emits a digitized choice at the device level, rather than producing a continuum of analog activations to be thresholded and reduced downstream in electronics.
In this split pipeline, coherence is used to compute hypothesis scores in parallel (Secs.~\ref{sec:phase_delay} and \ref{sec:linear_lookup}), while nonlinearity is used primarily for digitization into a single selected address (Sec.~\ref{sec:wta}).

Prior polaritonic and photonic time-domain processors often rely on relaxation dynamics, time-delayed nonlinearities, or reservoir-style computation without requiring phase-coherent interference across the full decision window \cite{Tyszka2023LIFPolaritons,Mirek2021NanoLett}.
PWC targets a different primitive: it enforces wrap-free and coherence conditions (see Eqs.~\eqref{eq:wrap_free}--\eqref{eq:coherence} below) so that relative delays are represented as phases and compared by programmable linear interference, reserving nonlinearity mainly for reducing the score vector to a single index via mode competition.
A formal statement of LUT-SNN notation and the role of address selection within the overall layer computation is given in Supplement Sec.~S1, and Fig.~\ref{fig:arch} summarizes the computational abstraction and physical pipeline.


\noindent\textbf{Positioning.}
PWC implements a timing-native argmax: programmable interference produces $K$ hypothesis scores from relative latencies, and driven--dissipative competition returns a one-hot address.
The goal is a repeatable top-1 decision (an address), not an analog output.
This realizes the timing $\rightarrow$ index step directly in wave hardware; downstream computation can remain digital or neuromorphic.

\medskip
\noindent\textbf{Contributions.}
We make three main steps.
First, we formalize address selection from spike times as a hardware primitive and introduce a phase--delay isomorphism that enables timing-native comparison by coherent interference within a wrap-free decision window, together with coherence requirements for reliable operation (Sec.~\ref{sec:phase_delay}).
Second, we develop a programmable interferometric scoring stage and a driven--dissipative digitizer that performs competitive selection, and we place timing jitter, static phase disorder, and phase diffusion into a common phase-noise description that yields margin-based robustness and a boundary-first failure mode (Secs.~\ref{sec:linear_lookup}--\ref{sec:robustness}).
Third, we show how minimal timing primitives compose into small spike-time circuits and connect device-level misaddressing to task-level impact with a routing-limited mixture-of-experts benchmark (Secs.~\ref{sec:circuits} and \ref{subsec:attention_routing}); we also propose an intensity-only hardware-in-the-loop calibration protocol that tunes programmable phases using measurable output margins (Sec.~\ref{sec:calibration}).

At the linear level, the scoring stage can be viewed as a coherent correlation or matched-filter computation \cite{VanderLugt1964,Goodman2015StatisticalOptics}, but with phase-encoded spike times as the correlated variables and a discrete address produced only after nonlinear selection.
This differs from amplitude-encoded photonic inference engines aimed at analog MAC-style linear algebra \cite{Shen2017NatPhoton,Lin2018D2NN,Wetzstein2020Nature,Zhou2022PhotonicMatrixMult}, where the output of interest is typically a real-valued activation or analog correlation peak.
Relative to time-domain demonstrations that primarily use delayed feedback as dynamical memory or reservoir effects \cite{Topfer2020TimeDelayPolaritonic,Mirek2022TimeDelayedNN}, we enforce wrap-free/coherence conditions so that delays can be compared by programmable interference, and we reserve nonlinearity primarily for discretization into a single selected address.

Throughout, we assume a shared phase reference, operation within a wrap-free coherent window, and a readout stage capable of competitive selection.
We focus on hard top-1 indexing (one selected port/address) rather than soft weighting.
A platform-specific break-even analysis is beyond the scope of this work, as it would require direct measurements of the timestamping latency $t_{\mathrm{TDC}}(\delta t)$ and energy $E_{\mathrm{TDC}}(\delta t)$, the traffic associated with score formation, the argmax latency and energy as functions of $K$, as well as reset time and calibration cadence.
We therefore focus on the routing primitive itself and connect the measured misrouting probability to task-level performance impact 
(Sec.~\ref{subsec:attention_routing}).

The paper is organized as follows.
Sec.~\ref{sec:phase_delay} formalizes the address-selection primitive and introduces the phase--delay isomorphism and operating constraints.
Secs.~\ref{sec:linear_lookup} and \ref{sec:wta} develop the linear interferometric lookup and the nonlinear WTA digitizer.
Sec.~\ref{sec:circuits} builds composable timing-native primitives and small cascades, and Secs.~\ref{sec:robustness} and \ref{sec:calibration} analyze robustness, scaling and calibration.
Sec.~\ref{sec:platforms_outlook} discusses candidate platforms, feasibility, and systems connections.
Supporting derivations appear in Appendix~\ref{app:analytics}; additional LUT-SNN definitions, statistics, simulations, calibration details, and feasibility analyses are provided in the Supplemental Material \cite{sm}.

\begin{figure*}[t]
  \centering
  \includegraphics[width=\textwidth]{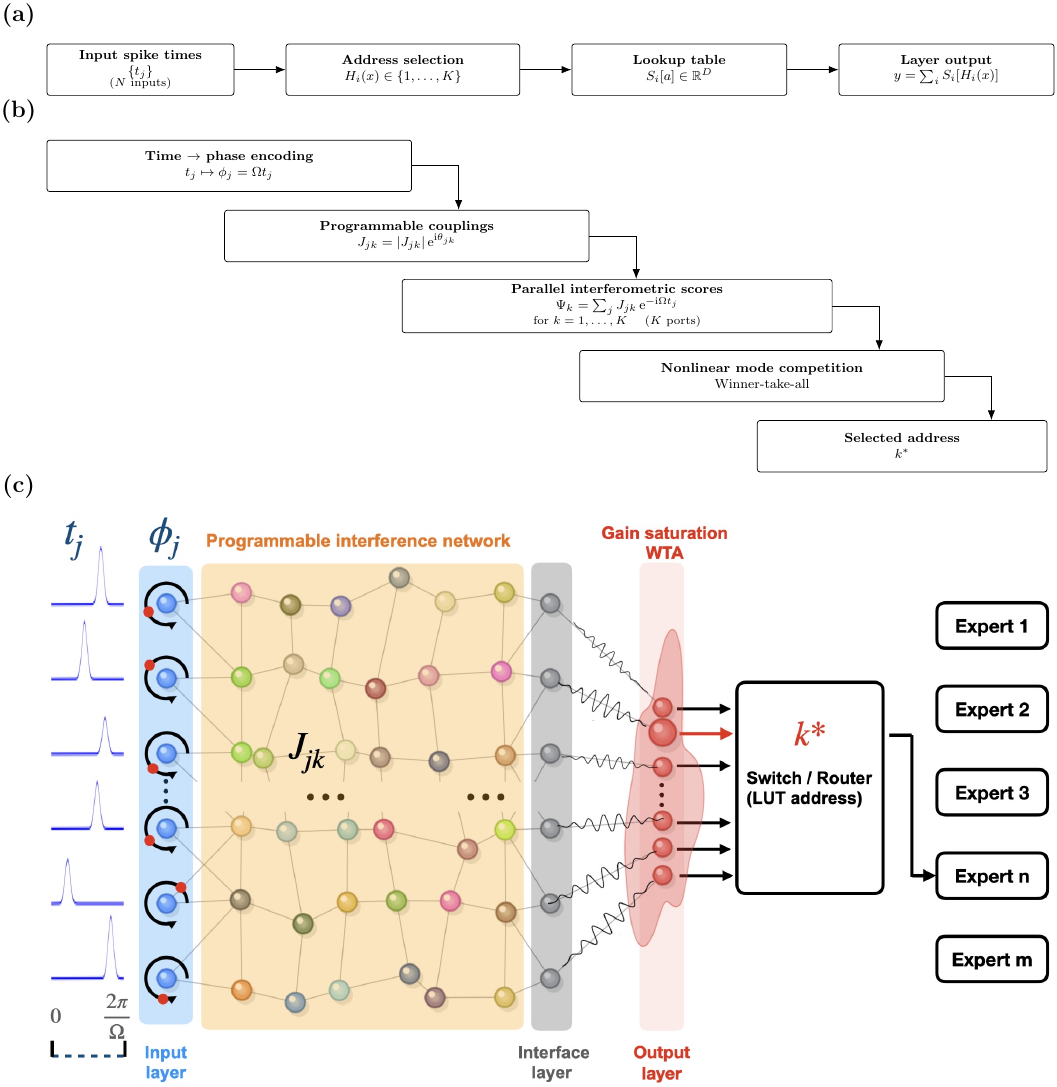}
  \caption{\textbf{Polychronous wave computing for LUT-style spiking.}
(a) Computational abstraction: a LUT-SNN layer maps spike times $\{t_j\}_{j=1}^{N}$ (latency vector $x$) to discrete addresses $H_i(x)\in\{1,\ldots,K\}$, retrieves synaptic vectors $S_i[H_i(x)]\in\mathbb{R}^{D}$, and accumulates $y=\sum_i S_i[H_i(x)]$.
(b) Physical realization: spike times are encoded as phases $\phi_j=\Omega t_j$ ($\Omega$ is an engineered reference (beat) frequency) within a wrap-free window $T_{\mathrm{wrap}}=2\pi/\Omega$; programmable complex couplings $J_{jk}=|J_{jk}|\mathrm{e}^{\mathrm{i}\theta_{jk}}$ evaluate all $K$ candidates by interference to produce amplitudes $\{\Psi_k\}$; driven--dissipative mode competition implements WTA selection, yielding a single discrete address $k^\ast$.
(c) Device-system-level cartoon: temporally encoded inputs drive a programmable interference network whose outputs feed an output-node layer embedded in a shared reservoir (shaded red irregular region); nonlinear competition selects a single winning output, which is forwarded to a downstream switch (router) that activates one selected expert (hard MoE routing). Wave motifs indicate coherent pulses and relative phase accumulation; they are schematic and do not represent resolved optical oscillations.}

  \label{fig:arch}
\end{figure*}

\section{Polychronous wave computing and phase--delay isomorphism}
\label{sec:phase_delay}

This section defines the address-selection operation and the phase--delay encoding that allows evaluation by coherent interference.

\subsection{Address selection from spike times}
\label{subsec:problem_statement}

Given spike times $x=(t_1,\ldots,t_N)$ from $N$ participating input channels (one spike per channel within the decision window), the device outputs one of $K$ discrete addresses
\begin{equation}
H:\mathbb{R}^N\to\{1,\ldots,K\}.
\label{eq:address_map_def}
\end{equation}
Here 
$j$ labels a fixed input channel (synapse); $t_{j}$
is the event time on that channel. The vector 
$x$ is not assumed sorted, and 
$j$ does not denote temporal order. The map $H$ partitions timing space into buckets $B_k=\{x:H(x)=k\}$; downstream computation depends on the selected index, not on the exact timing within a bucket.

A physical address selector should provide:
\begin{enumerate}
\item \textbf{Discrete output.} Return a single index $k\in\{1,\ldots,K\}$ rather than an analog score vector.
\item \textbf{Native-time evaluation.} Run at the platform timescale (ps--ns in ultrafast media) without serializing through timestamp conversion and clocked comparison.
\item \textbf{Robustness.} Small perturbations $x\mapsto x+\delta x$ should not change $H(x)$ except near bucket boundaries.
\end{enumerate}

Conventional implementations digitize each $t_j$ (timestamps, bins), compute similarity scores, and apply an electronic $\arg\max$.
Here the comparison is performed in the wave domain: spike times are encoded relative to a phase reference, candidate addresses are scored by programmable interference, and a driven--dissipative mode-competition stage produces a WTA address.

\subsection{Delay-to-phase mapping in a rotating frame}
\label{subsec:delay_phase_mapping}

Consider $N$ spike events at unordered times $\{t_j\}_{j=1}^N$ within a decision window
$
0 \le t_j \le t_{\max}.
$
Throughout we consider a latency code with one spike per participating input channel per decision window. Thus $N$ denotes the number of spikes (and participating channels) in that window.
Each spike triggers (or gates) a short coherent excitation derived from a common phase reference at angular frequency $\Omega$. Here $\Omega$ is an effective phase-reference frequency in the rotating frame (e.g., a beat note, mode splitting, or modulation tone), not necessarily the optical carrier.
Equivalently, each spike can be viewed as launching a narrowband coherent carrier whose complex envelope persists until a common readout time $T$.
At readout the envelope has accumulated phase $e^{i\Omega (T-t_j)}$ in the rotating frame, so the spike time is stored as a phase offset relative to the same reference.
A multiport interferometer applied at readout then combines these stored envelopes.
In a narrowband coherent system, a delay $t_j$ corresponds to a rotating-frame phase offset
\begin{equation}
\phi_j \equiv \Omega (t_j - t_{\rm ref})\ (\mathrm{mod}\ 2\pi),
\label{eq:phi_def}
\end{equation}
where $t_{\rm ref}$ is a reference time within the window. We take $t_{\rm ref}$ as the window start and set $t_{\rm ref}=0$.
Throughout, we represent the phase offset by the unit phasor
\begin{equation}
u_j \equiv e^{-i\phi_j},
\label{eq:phasor_def}
\end{equation}
so that increasing $t_j$ rotates $u_j$ clockwise on the unit circle (by our sign convention).

Relative delays map to relative phases,
\begin{equation}
t_j-t_m \;\longleftrightarrow\; \frac{\phi_j-\phi_m}{\Omega},
\label{eq:rel_delay_rel_phase}
\end{equation}
so a spike-time pattern becomes a set of unit phasors that can be compared by interference.

The linear score used in the lookup stage is a coherent sum evaluated at readout; it does not require temporal overlap of all spikes.
Each spike writes a complex contribution when it occurs, and the device samples the accumulated complex envelope at a common readout time $T$ near the end of the window.

A minimal model is a leaky coherent accumulator for each template (output) channel $k$.
Let $x\equiv(t_1,\ldots,t_N)$ denote the input latency pattern in the current decision window, and let
$\Psi_k(t;x)\in\mathbb{C}$ be the complex envelope (rotating-frame field amplitude) accumulated on channel $k$
at physical time $t$ in response to that pattern. We model $\Psi_k(t;x)$ as
\begin{equation}
\frac{\partial{\Psi}_k(t;x)}{\partial t}= -\left(\frac{1}{\tau}- i\Omega\right)\Psi_k(t;x) + \sum_{j} J_{jk}\,\delta(t-t_j),
\label{eq:accumulator}
\end{equation}
where $\tau$ is the complex-envelope lifetime in the rotating frame and the $\delta$-kicks represent event-triggered writes
of complex amplitude weighted by the programmed coefficients $J_{jk}$.
Solving Eq.~\eqref{eq:accumulator} and sampling at time $T$ gives
\begin{equation}
\Psi_k(T;x)=\sum_{j} J_{jk}\,e^{-(T-t_j)/\tau}\,e^{i\Omega(T-t_j)}.
\label{eq:accumulator_solution}
\end{equation}

For $\tau \gtrsim t_{\max}$ the decay factor $e^{-(T-t_j)/\tau}$ varies weakly across the decision window.
Define $\rho \equiv  t_{\max}/\tau$ and write $\Psi_k(T;x)=\sum_j J_{jk}\,w_j\,e^{i\Omega(T-t_j)}$ with $w_j \equiv e^{-(T-t_j)/\tau}$.
For readout times $T$ within (or near the end of) the window, $T-t_j\in[0, t_{\max}]$ and hence $w_j\in[e^{-\rho},1]$.

Let $\Psi_k^{(0)}(T;x)\equiv\sum_j J_{jk}\,e^{i\Omega(T-t_j)}$ denote the corresponding no-decay score.
Using the triangle inequality,
\begin{align}
\bigl|\Psi_k(T;x)&-\Psi_k^{(0)}(T;x)\bigr|
= \left|\sum_j J_{jk}\,(w_j-1)\,e^{i\Omega(T-t_j)}\right|\nonumber \\
& \le \sum_j |J_{jk}|\,|w_j-1|  \nonumber\\
&\le (1-e^{-\rho})\sum_j |J_{jk}|
\;\approx\; \rho\sum_j |J_{jk}|, 
\label{eq:decay_bound}
\end{align}
for $\rho\ll 1$.
Neglecting decay therefore perturbs the complex score by $O(\rho)$ (aside from an overall scale) and is appropriate when the bound in Eq. \ref{eq:decay_bound} is small compared with the winner–runner-up separation.

Up to a global phase $e^{i\Omega T}$ common to all $k$, Eq.~\eqref{eq:accumulator_solution} reduces to a phasor-sum form,
\begin{equation}
\Psi_k(T;x)\ \propto\ \sum_j J_{jk}\,e^{-i\Omega t_j},
\end{equation}
which is the scoring rule used throughout the linear lookup stage.
A global time shift $t_j\mapsto t_j+\delta t$ multiplies each input phasor by the same phase, so $\Psi_k\mapsto e^{-i\Omega\delta t}\Psi_k$, making the 
measurable intensities $I_k^{(\mathrm{lin})}=|\Psi_k|^2$ to be invariant to such shifts. In the following, we denote the (linear) interferometric score used for lookup as the readout-time value
$
\Psi_k(x)\equiv \Psi_k(T;x),
$
with a fixed readout time $T$ near the end of the window.

\subsection{Operating constraints for phase-coded temporal lookup}
\label{subsec:operating_constraints}

The encoding in Eq.~\eqref{eq:phi_def} is modulo $2\pi$, so distinct delays alias once the decision window spans more than one period.
To preserve a one-to-one delay-to-phase representation over the window we require
\begin{equation}
 t_{\max} < T_{\mathrm{wrap}} \equiv \frac{2\pi}{\Omega}.
\label{eq:wrap_free}
\end{equation}
If Eq.~\eqref{eq:wrap_free}  is violated, spikes separated by $T_{\mathrm{wrap}}$ are indistinguishable to any
phase-sensitive interferometric processor that uses only the single-tone mapping
$t \mapsto \Omega t\ (\mathrm{mod}\ 2\pi)$; this aliasing cannot be corrected by downstream WTA selection.

Interferometric evaluation also requires mutual coherence across the full delay span.
Let $\tau_{\mathrm{coh}}$ be the coherence time of the carrier field; a sufficient condition for phase-sensitive scoring is
\begin{equation}
\tau_{\mathrm{coh}} \gtrsim  t_{\max}.
\label{eq:coherence}
\end{equation}
This requirement is soft: finite coherence suppresses interference contrast continuously rather than producing a sharp failure, and can be folded into the phase-noise budget used for robustness analysis.

Timing jitter maps to phase noise.
If each spike time has rms jitter $\sigma_t$, then the corresponding rms phase uncertainty is
\begin{equation}
\sigma_{\phi} = \Omega\,\sigma_t.
\label{eq:jitter_phase}
\end{equation}
Equations~\eqref{eq:wrap_free}--\eqref{eq:jitter_phase} define the operating envelope: $\Omega$ and $ t_{\max}$ set the usable delay span via $T_{\mathrm{wrap}}$, while $\Omega\sigma_t$ sets the phase-noise penalty for temporal precision.

\section{Interferometric lookup and linear theory} \label{sec:linear_lookup} 

\subsection{Multiport interferometric scoring of spike-time patterns} \label{subsec:interferometric_scoring} Within the wrap-free coherent window of Sec.~\ref{sec:phase_delay}, an input spike at time $t_j$ is encoded as a unit-modulus phasor $u_j$ (Eq.~\ref{eq:phasor_def}). A spike pattern corresponds to a (typically sparse) vector $u\in\mathbb{C}^N$ whose phases encode timing.
We score the $K$ candidate templates using a linear $N\times K$ multiport interferometric network with programmable couplings
$J_{jk}=|J_{jk}|\,e^{i\theta_{jk}}$.
The complex field in output channel $k$ becomes
\begin{equation}
\Psi_k(x)
= \sum_{j=1}^{N} J_{jk}\, u_j
= \sum_{j=1}^{N} |J_{jk}|\exp\!\left[i\left(\theta_{jk}-\Omega t_j\right)\right].
\label{eq:psi_linear}
\end{equation}
Equation~(\ref{eq:psi_linear}) uses an effective input--output coupling matrix $J_{jk}$: the net complex transfer coefficient from input channel $j$ (a spike-launched phasor $u_j$) to output port (template) $k$ at the common readout time. The score is evaluated once at readout and represents the coherent sum of event-driven contributions accumulated over the decision window, producing $K$ template scores in parallel. In hardware the underlying coupling graph may be sparse and local (e.g., a polariton lattice \cite{Berloff2017NatMater}), but multi-path propagation through the programmable interference region and the associated phase accumulation can yield a dense effective $J_{jk}$.

Most platforms measure power; we therefore define the linear intensity score
\begin{equation}
I^{(\rm lin)}_k(x) \equiv |\Psi_k(x)|^2,
\label{eq:score_intensity}
\end{equation}
which provides the seed for the nonlinear WTA readout in Sec.~\ref{sec:wta}.

At the linear level, Eq.~\eqref{eq:psi_linear} is a coherent matched-filter computation \cite{VanderLugt1964,Goodman2015StatisticalOptics}; here the correlated variables are phase-encoded spike times and the output of interest is a discrete address after nonlinear selection.

\subsection{Matched channel, unmatched background, and margin metrics}
\label{subsec:matched_unmatched}

A template $k$ is specified by reference spike times $\{t_j^{(k)}\}_{j=1}^N$ within the decision window (equivalently by phases $\Omega t_j^{(k)}$).
Compiling the interferometer to that template corresponds to choosing
\begin{equation}
\theta_{jk}=\Omega t_j^{(k)},
\label{eq:matched_phase}
\end{equation}
so that the linear score at port $k$ for an input $x=(t_1,\dots,t_N)$ is
\begin{equation}
\Psi_k(x) = \sum_{j=1}^{N}|J_{jk}|
\exp\!\left[i\Omega\left(t_j^{(k)}-t_j\right)\right].
\label{eq:matched_sum}
\end{equation}
For a perfect match $k=k^\ast$ such that $t_j=t_j^{(k^\ast)}$ for all $j$, the terms add in phase and the matched amplitudes scale as 
\begin{equation}
A_{k^\ast}(x)\equiv|\Psi_{k^\ast}(x)| \approx \sum_{j=1}^{N}|J_{jk^\ast}|
\sim N\,\bar{J},
\label{eq:matched_scaling}
\end{equation}
where $\bar J$ denotes a typical coupling magnitude (e.g., the mean of $|J_{j k^\ast}|$).
More explicitly, the matched amplitude is
$A_{k^\ast}(x)=\sum_{j=1}^N |J_{j k^\ast}|$, which is $O(N)$ provided no single term dominates the sum.

\paragraph{Unmatched background (random-phase null).}
For $k\neq k^\ast$, we treat the relative phases $\theta_{jk}-\Omega t_j$ as approximately independent and uniform on $[0,2\pi)$.
Then $\mathbb{E}[\Psi_k]\simeq 0$ and
\begin{equation}
\mathbb{E}[|\Psi_k|^2]\simeq \sum_{j=1}^{N}|J_{jk}|^2,
\label{eq:null_second_moment}
\end{equation}
so a rms unmatched amplitude scales as
\begin{equation}
A_{k\ne k^\ast}\equiv |\Psi_{k\ne k^\ast}|
\sim \bar{J}\sqrt{N}
\quad
\text{(random-phase null)}.
\label{eq:unmatched_scaling}
\end{equation}
This gives the  separation
\begin{equation}
A_{k^\ast}=\mathcal{O}(N),
\qquad
A_{k\ne k^\ast}=\mathcal{O}(\sqrt{N}).
\label{eq:match_mismatch_scaling}
\end{equation}
This $O(N)$ versus $O(\sqrt{N})$ separation sets the typical margin scale for addressing.

\paragraph{Linear discrimination margin (ground-truth).}
Addressing depends on the separation between the correct channel and the strongest incorrect competitor.
We define the ground-truth amplitude margin as
\begin{equation}
\Delta_{\mathrm{lin}}(x) \equiv A_{k^\ast}(x)-\max_{k\neq k^\ast}A_k(x),
\label{eq:Delta_lin_def}
\end{equation}
and the log-intensity margin used in Fig.~\ref{fig:linear_lookup} as
\begin{equation}
\Delta_{\mathrm{lin}}^{\log}(x)\equiv
\log\!\left(\frac{I_{k^\ast}^{(\mathrm{lin})}(x)}{\max_{k\neq k^\ast} I_k^{(\mathrm{lin})}(x)}\right),
\label{eq:Delta_lin_log_def}
\end{equation}
where $\log$ stands for the natural logarithm.
These margins share the same decision boundary:
correct linear addressing is $\Delta_{\mathrm{lin}}(x)>0$ (equivalently $\Delta_{\mathrm{lin}}^{\log}(x)>0$).
If the matched channel is not the linear leader, then any deterministic WTA stage seeded by $\{\Psi_k\}$ cannot return the correct index, since the ordering is already incorrect.

For calibration and experiments, we use directly measurable post-readout observables $I_k(x)$ (steady-state intensities or time-integrated energies) and define the intensity margin
\begin{equation}
M(x)\equiv I_{k^\ast}(x)-\max_{k\neq k^\ast} I_k(x),
\label{eq:M_def}
\end{equation}
which is the objective in the hardware-in-the-loop protocol (Sec.~\ref{sec:calibration}).
For linear readout,
\begin{eqnarray}
M_{\mathrm{lin}}(x)&=& I_{k^\ast}^{(\mathrm{lin})}(x)-\max_{k\neq k^\ast} I_k^{(\mathrm{lin})}(x)
= A_{k^\ast}^2-\max_{k\neq k^\ast}A_k(x)^2\nonumber \\
&=& \Delta_{\mathrm{lin}}(x)\,\bigl(A_{k^\ast}(x)+\max_{k\neq k^\ast}A_k(x)\bigr).
\label{eq:Mlin_identity}
\end{eqnarray}
Equations~\eqref{eq:Delta_lin_def}, \eqref{eq:Delta_lin_log_def}, and \eqref{eq:M_def} define the margin conventions used throughout.

\subsection{Noise parameters and interferometric LUT discrimination.}
\label{subsec:linear_noise_bridge}

\begin{figure*}[t]
  \centering
  \includegraphics[width=\textwidth]{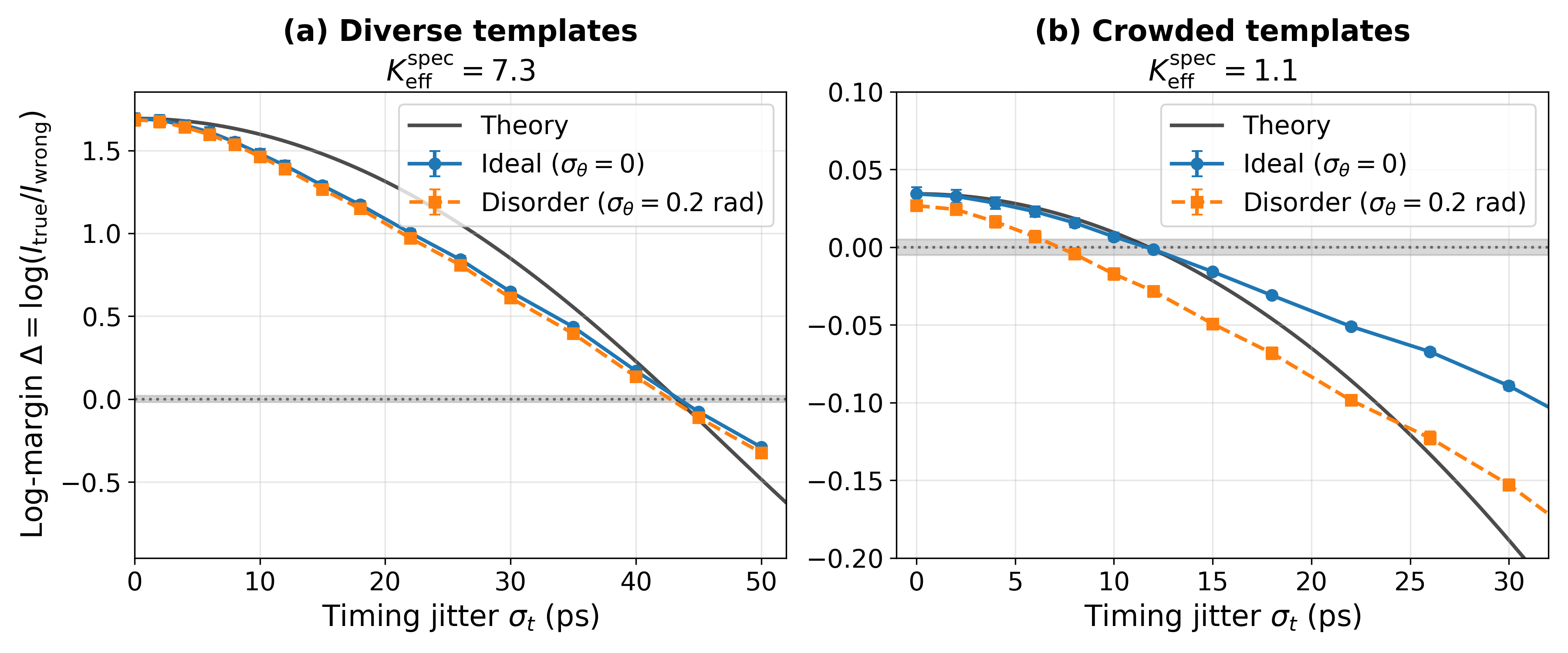}
\caption{\textbf{Robustness comparison of interferometric LUT discrimination.}
Linear log-margin $\Delta=\log\!\big(I_{\mathrm{true}}/I_{\mathrm{wrong}}\big)$ versus timing jitter $\sigma_t$
for $K=8$ templates and $N=32$ input spikes; the gray band marks the decision boundary $\Delta=0$.
Blue/orange points show Monte Carlo means $\pm$ SEM over $5$ independent template banks and $200$ trials per class
(ideal interferometer $\sigma_\theta=0$; static per-coupling phase disorder $\sigma_\theta=0.2~\mathrm{rad}$).
Solid black curves show a semi-analytic coherence-decay model for the log-margin under phase jitter
$\sigma_\phi=\Omega\sigma_t$.
Using $\mathbb{E}[I_{\rm true}]=N+N(N-1)e^{-\sigma_\phi^2}$ and absorbing the competitor baseline into a fitted offset,
we plot
$\Delta_{\rm th}(\sigma_t)=\log\!\big(1+(N-1)e^{-\alpha\sigma_\phi^2}\big)-\log N+\Delta_0$,
with $(\alpha,\Delta_0)$ fitted separately for the diverse and crowded regimes.
(a) \emph{Diverse} templates (i.i.d.\ uniform times; $K_{\mathrm{eff}}^{\mathrm{spec}}=7.25\pm0.11$): disorder has negligible effect and the
zero-crossing remains near $\sigma_t\simeq 43~\mathrm{ps}$.
(b) \emph{Crowded} templates (sorted times within each pattern; $K_{\mathrm{eff}}^{\mathrm{spec}}=1.13\pm 0.03$): disorder shifts the
zero-crossing from $\sim11.6~\mathrm{ps}$ to $\sim7.2~\mathrm{ps}$ (a $4.4~\mathrm{ps}$ operating-window reduction). Template diversity, quantified by $K_{\mathrm{eff}}^{\mathrm{spec}}$, acts as a robustness knob.
}
\label{fig:linear_lookup}
\end{figure*}

Timing jitter $t_j\mapsto t_j+\delta t_j$ maps to phase perturbations $\delta\phi_j=\Omega\,\delta t_j$ (so $\sigma_\phi=\Omega\sigma_t$),
while static phase offsets in the compiled couplings act as additional phase errors.
In the linear lookup stage, these imperfections suppress coherent contrast and increase the likelihood that the nearest competitor overtakes the leader, so errors concentrate near winner--runner-up near-ties.
Figure~\ref{fig:linear_lookup} summarizes this behavior under timing jitter and static phase disorder and
compares simulations to a semi-analytic coherence-decay model for the log-margin (black curves).
A unified phase-noise budget and competitor scaling are developed in Sec.~\ref{sec:robustness} (with derivations in Appendix~\ref{app:jitter_disorder} and Appendix~\ref{app:extreme_value}).
Disorder sensitivity depends strongly on library crowding because the zero-jitter margin can differ by orders of magnitude. In diverse libraries the margin is large, so $\sigma_\theta=0.2$ rad has little effect; in deliberately crowded libraries the margin lies near the boundary and the same disorder reduces the operating window by several picoseconds.

In both cases the margin degrades smoothly with increasing jitter rather than failing abruptly, motivating nonlinear WTA readout, which converts finite linear margins into robust discrete address selection (Sec.~\ref{sec:wta}).

\section{Nonlinear mode competition and winner-take-all selection}
\label{sec:wta}

The interferometric stage provides complex seeds $\Psi_k$ [Eq.~\eqref{eq:psi_linear}] that quantify linear template matches (Sec.~\ref{sec:linear_lookup}).
Lookup-style computation requires a discrete index rather than an analog score vector.
Driven--dissipative systems can perform this analog-to-digital step by mode competition: above threshold, modes draw from a limited gain resource and gain saturation allows one mode to reach macroscopic occupation while suppressing the rest \cite{Maass2000WTApower,wouters2007excitations,Carusotto2013RMP,Kalinin2019PolaritonicNetwork}.

\subsection{Physical digitization as a driven--dissipative \texorpdfstring{$\arg\max$}{argmax}}
\label{subsec:physical_argmax}

Given $\{\Psi_k\}$, the nonlinear readout selects a single index $k^\ast$.
The readout acts as a physical argmax: the strongest-seeded channel reaches threshold first and clamps the gain.
This differs from polaritonic neuromorphic approaches where nonlinearity is used mainly as an analog activation or memory element \cite{ballarini2020neuromorphic,Mirek2021NanoLett,Tyszka2023LIFPolaritons}.
Here the goal is a reproducible discrete decision: map the continuous contrasts ${|\Psi_k|}$ to a single winning index.

\subsection{Microscopic description and compilation of interferometric seeds}
\label{subsec:odgpe_and_seeding}

We separate (i) coherent phase-sensitive scoring (kept linear to preserve interference) from (ii) nonlinear digitization into a single winner (via gain competition).
To anchor the discussion in a polariton implementation while keeping the selection principle general, we start from the open-dissipative Gross--Pitaevskii (the complex Ginzburg--Landau) equation coupled to an incoherent reservoir   \cite{wouters2007excitations,Carusotto2013RMP,keeling2008spontaneous,Kalinin2019PolaritonicNetwork,Kalinin2020XYIsing}:
\begin{align}
i\hbar \frac{d\psi_k}{dt} &=
\left[
\epsilon_k
+ g_c |\psi_k|^2
+ g_R n_k
+\frac{i\hbar}{2}\left(R n_k-\gamma_c\right)
\right]\psi_k\nonumber \\
&+ \sum_{\ell\neq k} J_{k\ell}\,\psi_\ell
+ F_k(t),
\label{eq:odgpe_main}
\\
\frac{dn_k}{dt} &= P_k(t)-\left(\gamma_R+R|\psi_k|^2\right)n_k .
\label{eq:reservoir_main}
\end{align}
Here $\psi_k$ is the condensate field for channel $k$ and $n_k$ is the associated reservoir density.
$\gamma_c,\gamma_R$ are decay rates, $\epsilon_k$ is the on-site detuning, $R$ is the stimulated scattering rate, and $g_c,g_R$ parameterize interaction-induced shifts.
The coherent couplings $J_{k\ell}=|J_{k\ell}|e^{i\theta_{k\ell}}$ describe mode mixing.

The coherent term $F_k(t)$ is a brief injection at readout onset with complex amplitude proportional to the upstream interferometric score $\Psi_k$, so relative seeds encode score contrast.  

It is sufficient that the injection be short compared to the subsequent selection dynamics, so it acts effectively as an initial condition.
The incoherent pump $P_k(t)$ sets the reservoir operating point during readout and enables gain competition and reset; pulsed reservoir dynamics can realize leaky integration, thresholding, and reset on sub-ns timescales \cite{Tyszka2023LIFPolaritons,Opala2023HarnessingPolaritons}.

\subsection{Reduced competition model and conditions for single-winner selection}
\label{subsec:reduced_wta}

Equations~\eqref{eq:odgpe_main}--\eqref{eq:reservoir_main} mix coherent evolution/mode coupling, gain and depletion, and interaction-induced shifts.
For analysis it is convenient to use reduced models aligned with the pipeline: a coherent model for timing-to-routing contrast, and a competition model for digitization.

\paragraph{Digitization via gain competition (envelope WTA model).}
Near threshold, driven--dissipative oscillators admit Stuart--Landau-type reductions with saturable gain
\cite{Kalinin2019PolaritonicNetwork}.
We model readout competition by complex envelopes $\{\psi_k(t)\}_{k=1}^K$ with self- and cross-saturation:
\begin{equation}
\frac{d \psi_k}{dt}=
\Bigg[
(G_k-\gamma)
-\eta |\psi_k|^2
-\sum_{\ell\neq k}\chi_{\ell k}\,|\psi_\ell|^2
\Bigg]\psi_k
+g_{\rm inj}\,\Psi_k .
\label{eq:dd_wta_revised}
\end{equation}
Here $G_k$ is the small-signal gain, $\gamma$ the amplitude loss rate, $\eta>0$ the self-saturation coefficient,
and $\chi_{\ell k}\ge 0$ the cross-saturation due to a shared gain resource.
The additive drive $g_{\rm inj}\,\Psi_k$ seeds each mode in proportion to the upstream interferometric score $\Psi_k$
[Eq.~(\ref{eq:psi_linear})], matching the reduced competitive digitizer used in Supplement Sec.~3.2 [Eq.~(S22)]
(where $G_k\!=\!G$ and $\chi_{\ell k}\!=\!\chi$ are taken uniform).

Because winner selection depends only on intensities $|\psi_k|^2$, we may take $g_{\rm inj}\ge 0$ real without loss of generality:
any constant phase factor in the complex seed can be absorbed into the definition of $\psi_k$.
Equivalently, if one prefers the wave-model forcing convention $\dot\psi=\cdots - iF(t)$,
the same seeding can be written as $-iF_k$ with $F_k \equiv i\,g_{\rm inj}\,\Psi_k$.

In a physical device the seed can be implemented as a brief injection or, equivalently, as an initial-condition bias;
once the injection is removed, the subsequent autonomous gain-competition dynamics are those analyzed below in
Eqs.~(\ref{eq:intensity_competition})--(\ref{eq:address_readout_revised}).
Equation~\eqref{eq:dd_wta_revised} is used to quantify single-winner reliability under noise and disorder given seed contrasts
(e.g.\ Fig.~\ref{fig:wta_hero} and 
Supplement Sec.S3).

\paragraph{Coherent routing/scoring in compact junctions (wave model).}
For temporal primitives and small cascades, we model the interferometric routing dynamics explicitly at the wave level,
focusing on the coherent conversion of spike timing into output-port contrast.
We simulate a coupled-mode envelope model
\begin{equation}
\frac{d\boldsymbol{\psi}}{dt}
=
-i\,{\cal H}\,\boldsymbol{\psi}
-i\,U\!\left(|\boldsymbol{\psi}|^2\odot \boldsymbol{\psi}\right)
-\gamma\,\boldsymbol{\psi}
-i\,\mathbf{F}(t),
\label{eq:complex_envelope_main}
\end{equation}
with $\boldsymbol{\psi}=(\psi_1,\ldots,\psi_K)^\mathsf{T}$ (or $(\psi_A,\psi_B,\psi_L,\psi_R)^\mathsf{T}$ for the four-mode comparator).
Here ${\cal H}$ encodes the linear detunings and couplings of the interferometric network, $\gamma$ is the loss rate,
$U$ is a Kerr nonlinearity, and $\mathbf{F}(t)$ represents time-dependent coherent injection corresponding to input spikes.
The notation $\odot$ denotes elementwise multiplication.

This wave-level model treats interferometric routing directly by integrating the driven coupled-mode dynamics
\eqref{eq:complex_envelope_main} and reading out port contrasts from $\boldsymbol{\psi}(t)$.
Elsewhere we use an equivalent reduced description in terms of an effective input--output transfer matrix $J_{jk}$,
defined as the readout-time complex coefficient mapping an impulsive excitation at input port $j$ to the field at
output port (template) $k$ with the same normalization and envelope-decay conventions as in Sec.~\ref{sec:linear_lookup}.
In Figs.~\ref{fig:primitives}--\ref{fig:N3_LUT} we set $U=0$ to isolate linear interference; digitization is handled downstream
by Eq.~\eqref{eq:dd_wta_revised} or the reservoir model. Full Hamiltonians, pulse definitions, readout windows, and numerical
parameters are given in the Supplemental Material.

\paragraph{Relation between the wave model and the compiled lookup couplings.}
The operator ${\cal H}$ encodes the physical interferometric network (detunings and couplings) in the wave model, and $J_{jk}$
is the corresponding induced transfer coefficient at readout (not an additional parameter).
Equivalently, $J_{jk}$ can be viewed as the appropriate Green's-function element of ${\cal H}$ sampled at the readout time. For the linear, time-invariant case ($U=0$), define the generator
$A \equiv -iH - \gamma I$ so that $\dot\psi = A\psi - iF(t)$, with Green's function
$G(\Delta t)=e^{A\Delta t}$.
An impulsive excitation at input $j$ at time $t_j$ contributes a term proportional to
$G_{k j}(T-t_j)$ at output port $k$ at readout time $T$ (up to convolution with the pulse envelope).
The compiled coefficients $J_{jk}$ used in Eq.~\ref{eq:matched_phase} summarize this effective readout-time transfer
under the conventions of Secs.~\ref{sec:phase_delay} -- \ref{sec:linear_lookup}.

It  has been  shown experimentally that a spatially programmable multimode waveguide can be trained to realize an arbitrary linear input–output map by shaping the refractive-index landscape of a continuous medium. In that setting, the trained wave propagation defines an effective linear operator between input and output fields at a fixed readout plane \cite{Onodera2025MultimodeWaveTraining}. Our 
$J_{jk}$
 plays an analogous role, but is explicitly discretized over labeled input ports and template outputs and is optimized for phase-encoded timing discrimination rather than amplitude regression.

\paragraph{Single-winner conditions (after seeding).}
Writing $I_k\equiv|\psi_k|^2$ and considering times after injection ($F_k\to 0$) in the digitizer limit (${\cal H}_{k\ell}=0$) gives
\begin{equation}
\frac{d I_k}{dt}
=
2\left[
(G_k-\gamma)
-\eta I_k
-\sum_{\ell\neq k}\chi_{\ell k}\,I_\ell
\right] I_k .
\label{eq:intensity_competition}
\end{equation}
A single-winner fixed point has the form
\begin{equation}
I_{k^\ast}\approx\frac{G_{k^\ast}-\gamma}{\eta},
\qquad
I_{\ell\neq k^\ast}\to 0 ,
\label{eq:wta_fixed_point_revised}
\end{equation}
provided cross-saturation destabilizes multimode coexistence (e.g.\ for two symmetric modes, coexistence becomes unstable when $\chi=\chi_{12}>\eta$).
The discrete address is read out as
\begin{equation}
H(x)=\arg\max_k I_k(t_{\rm read}),
\label{eq:address_readout_revised}
\end{equation}
with $t_{\rm read}$ after saturation separates the winner.
If residual coherent coupling remains significant during readout (${\cal H}_{k\ell}\neq 0$),
instantaneous intensities $I_k(t)=|\psi_k(t)|^2$ can exhibit beating and the instantaneous winner can
depend on the sampling time. In that regime we define the readout score as the time-integrated energy
$$
E_k \equiv \int_{t_{\mathrm{read}}}^{t_{\mathrm{read}}+T_w} |\psi_k(t)|^2\,dt,
$$
and report $H(x)=\arg\max_k E_k$.
This agrees with the post-saturation winner when $T_w$ spans several beat periods and/or the saturation
rate exceeds $\|H\|$ (Supplement Sec.~S4; see also Sec.~S5).


\subsection{Margin as a label-free confidence proxy}
\label{subsec:address_map_and_margin}

The WTA readout implements the address map $H$ introduced in Sec.~\ref{subsec:problem_statement},
with $\hat{k}=H(x)=\arg\max_k I_k(t_{\rm read})$.
To quantify how close a query is to a decision boundary using only the interferometric scores,
we use the winner--runner-up gap
\begin{equation}
\Delta_{\mathrm{win}}(x)\equiv |\Psi_{\hat{k}}(x)|-\max_{k\neq \hat{k}}|\Psi_k(x)|.
\label{eq:margin_revised}
\end{equation}
Unlike the ground-truth margin $\Delta_{\rm lin}$ (Sec.~\ref{sec:linear_lookup}), $\Delta_{\mathrm{win}}$ is label-free and can be measured
at inference time as a confidence indicator. If all score magnitudes change by at most $\varepsilon$ under a perturbation,
then $\Delta_{\mathrm{win}}(x)>2\varepsilon$ is sufficient to preserve the winner. We use this bound only as an interpretable proxy;
all accuracies reported below are obtained from the full driven--dissipative readout dynamics.

\begin{figure*}[t]
  \centering
\includegraphics[width=\linewidth]{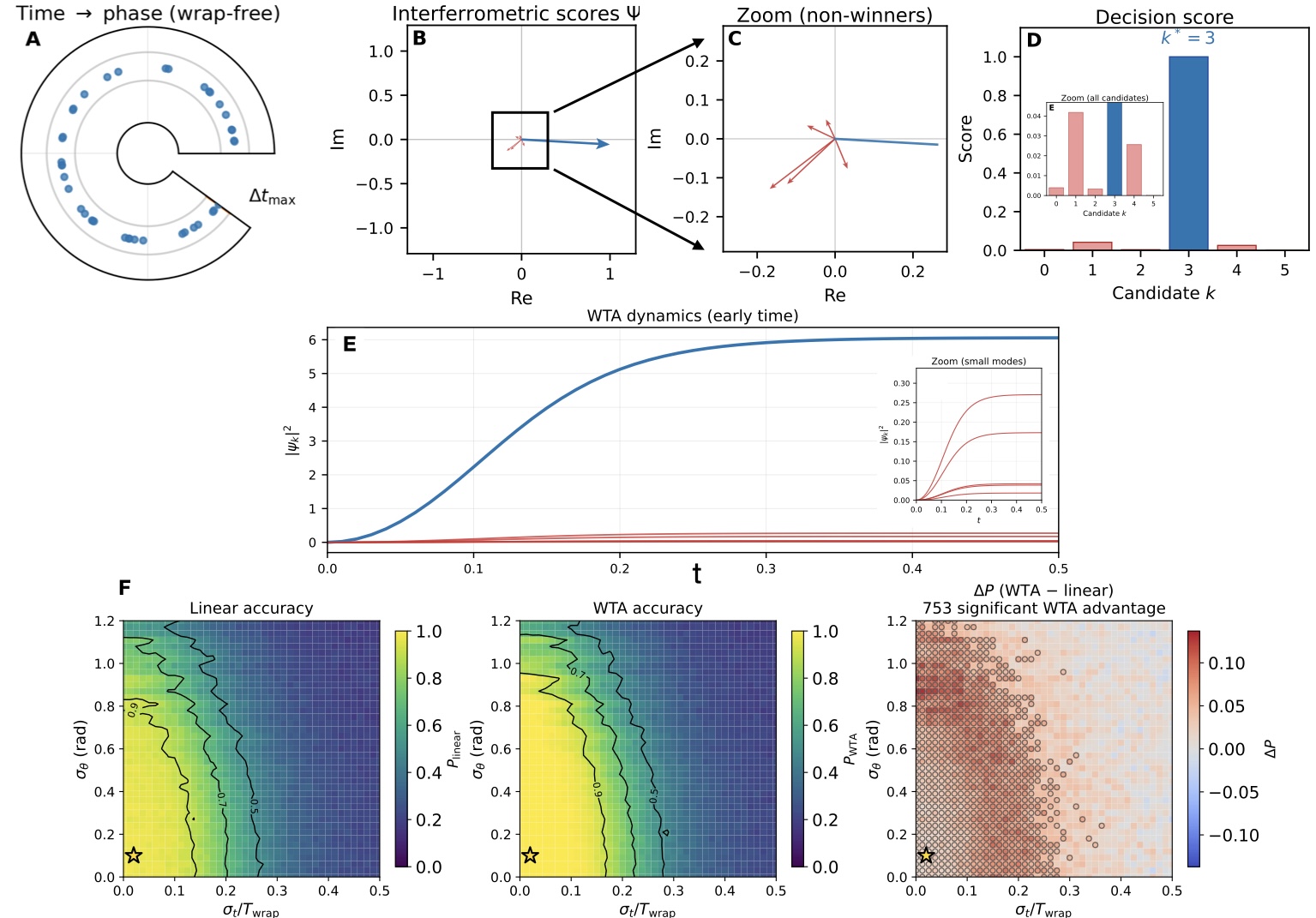}
\caption{\textbf{Multi-port nonlinear selection: scoring $\rightarrow$ seeded competition $\rightarrow$ discrete address.}
(\textbf{A}) \emph{Time-to-phase encoding (wrap-free).} A representative trial with $N=40$ input spikes mapped to phases
$\phi_j=\Omega t_j$ within a wrap-free window $\Delta t_{\max}<T_{\mathrm{wrap}}$.
(\textbf{B}) \emph{Interferometric template scores.} Complex scores for all $K=6$ candidates,
$\Psi_k=\sum_{j=1}^{N} J_{jk}e^{-i\Omega t_j}$ (with $J_{jk}=|J_{jk}|e^{i\theta_{jk}}$); the winner $k^\ast$ is highlighted.
(\textbf{C}) \emph{Zoom on non-winners.} Magnified view of the small-magnitude competing phasors near the origin.
(\textbf{D}) \emph{Decision score after digitization.} Bar plot of readout scores showing strong suppression of non-winners
(inset: zoom of all candidates); the reported address is $\hat{k}=\arg\max_k I_k(t_{\mathrm{read}})$.
(\textbf{E}) \emph{Early-time WTA dynamics.} Seeded gain competition amplifies the leading mode and suppresses the others,
showing the evolution of $I_k(t)=|\psi_k(t)|^2$ (inset: zoom of non-winner trajectories).
(\textbf{F}) \emph{Operating envelope.} Accuracy over timing jitter $\sigma_t/T_{\mathrm{wrap}}\in[0,0.5]$ and static phase disorder
$\sigma_\theta\in[0,1.2]$\,rad for $K=6$, $N=10$ on a $41\times41$ grid with $1000$ samples per pixel (trials $\times$ device realizations).
Left: linear readout accuracy $P_{\mathrm{linear}}$ from the seed intensities $I_k^{(\mathrm{lin})}=|\Psi_k|^2$
(with multiplicative readout noise); middle: WTA readout accuracy $P_{\mathrm{WTA}}$ from the winner of the driven--dissipative
competition model (Eq.~\eqref{eq:dd_wta_revised}); right: $\Delta P=P_{\mathrm{WTA}}-P_{\mathrm{linear}}$, with markers indicating
pixels where WTA provides a statistically significant advantage (criterion in the Supplemental Material). The star marks the operating point used in (\textbf{A}--\textbf{E}).}
\label{fig:wta_hero}
\end{figure*}

Perturbations from jitter, residual static phase error, and finite coherence act primarily by shrinking $\Delta_{\mathrm{win}}$,
so misaddressing onsets in near-tie events. Section~\ref{sec:robustness} quantifies this via an effective phase-noise budget and competitor statistics
(see also Supplemental Material Sec.~S3). Figure~\ref{fig:wta_hero} summarizes the scoring--digitization cascade.

\section{Primitive circuits and composability}
\label{sec:circuits}

WTA readout (Sec.~\ref{sec:wta}) converts interferometric contrasts into a one-hot outcome: one mode/port dominates while the rest are suppressed.
This yields the discrete symbol needed for lookup-style computation (an address or branch outcome) without explicit timestamping.

To move beyond single-shot addressing, we identify a small set of timing-native primitives and show how they compose.
This follows the lookup-table viewpoint \cite{Izhikevich2025SpikingManifesto}: compute an address from relative latencies, then retrieve a stored value.
Here the corresponding ingredients are realized in a coherent-wave substrate: temporal comparison, many-way scoring, and discrete selection (routing).

\subsection{Temporal order comparison as a one-bit primitive}
\label{subsec:order_primitive}

A minimal temporal operation is the order test between two spikes: given events $A$ and $B$ at times $t_A,t_B$, decide whether $t_A<t_B$ or $t_B<t_A$.
Pairwise order relations are central to polychronization \cite{Izhikevich2006Polychronization} and to LUT-style address construction via concatenated comparisons \cite{Izhikevich2025SpikingManifesto}.

\paragraph{Flux-biased interferometric comparator.}
A two-spike order test can be implemented by a small interferometric junction with two inputs $(A,B)$ and two readout ports $(L,R)$.
The first pulse seeds a coherent field and the second interferes with it.
With coupling phases chosen to enclose a nonzero synthetic gauge flux, the junction response becomes chiral: opposite temporal orders route energy to opposite outputs \cite{Fang2012EffectiveMagneticFieldPhotons,Ozawa2019TopologicalPhotonics}.
After the short linear interaction, WTA readout (Sec.~\ref{sec:wta}) digitizes the outcome using time-integrated output energies,
\begin{align}
b_{AB} &= \mathbbm{1}[E_L>E_R] \;=\; \mathbbm{1}[m_{AB}>0],
\\
m_{AB} &\equiv \log(E_L/E_R),
\end{align}
where $E_{L,R}$ are integrated intensities over a fixed window.
The bit $b_{AB}$ encodes the sign of $t_A-t_B$ and $m_{AB}$ provides a confidence margin. 
Because the sign is ultimately set by a sinusoidal dependence on the relative delay
(Appendix~\ref{app:order_comparator}, Eq.~(\ref{eq:II})), unambiguous order decoding requires the pairwise half-cycle constraint
\begin{equation}
|\Delta t|\equiv |t_A-t_B| < \pi/\Omega,
\label{eq:comparator_noalias_main}
\end{equation}
which is stricter than the global wrap-free window $t_{\max}<2\pi/\Omega$ in Eq.~\ref{eq:wrap_free}.
A minimal analytic model is given in Appendix~\ref{app:order_comparator}.

Figure~\ref{fig:primitives} shows the primitive in the linear-interference regime: opposite input orders produce opposite left/right imbalance, while the integrated-energy readout smooths transient beating and provides a signed analog confidence margin.
This margin is the primitive-level quantity that later controls composability: near-ties ($m_{AB}\approx 0$) are the only events that can flip the bit under jitter or disorder.

\begin{figure*}[t]
  \centering
  \includegraphics[width=0.92\textwidth]{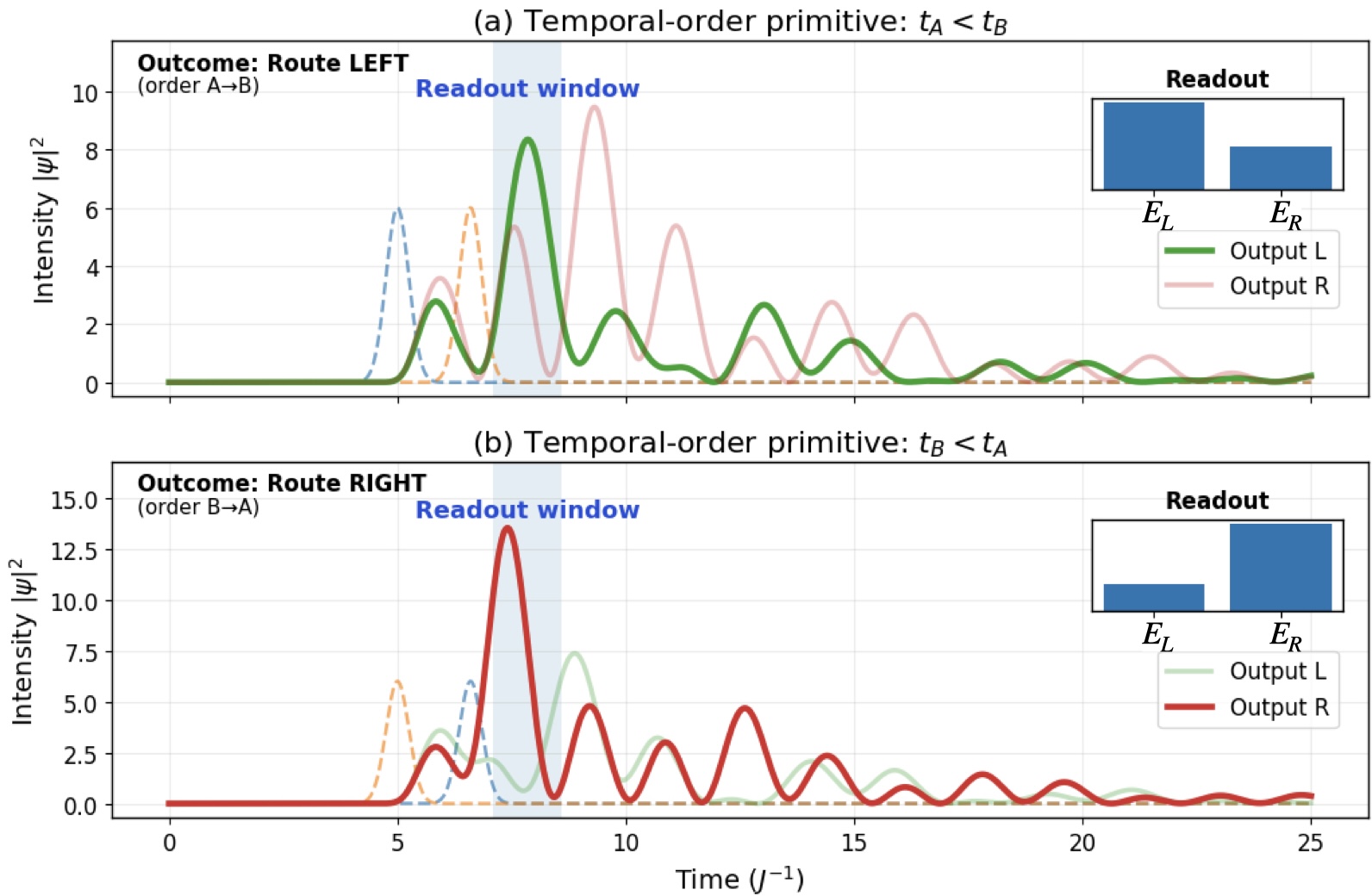}
  \caption{\textbf{Temporal-order comparator as a one-bit primitive.}
Two input pulses $A$ and $B$ excite a four-mode junction at times $t_A$ and $t_B$, producing left/right output intensities $|\psi_{L,R}(t)|^2$.
The junction routes energy predominantly left for $t_A<t_B$ (\textbf{a}) and right for $t_B<t_A$ (\textbf{b}).
The decision uses integrated readout energies $E_{L,R}=\int_{t_0}^{t_0+T}|\psi_{L,R}(t)|^2\,dt$ (shaded window),
yielding bit $b_{AB}=\mathbbm{1}[E_L>E_R]$ and margin $m_{AB}=\log(E_L/E_R)$.
The simulations use the linear regime of the four-mode model ($U=0$). Parameters and disorder realizations are in Supplement Sec.~S4.}
  \label{fig:primitives}
\end{figure*}

The comparator therefore implements a one-bit temporal hash with an analog reliability signal.
In the next subsection we use these bits as building blocks, and use the margins to detect and repair the rare inconsistency events that arise under noise.

\subsection{Routing and fan-out}
\label{subsec:routing_fanout}

Each primitive outputs a selected mode (port) rather than a multi-bit value.
In a wave system this directly specifies a spatial channel (mode/waveguide), so routing can be implemented by connectivity.
Because WTA suppresses competing modes, the representation is naturally one-hot; fan-out reduces to splitting the winning channel.
In practice, fan-out is limited by residual modal cross-talk and by how sharply WTA separates the winner from the runner-up, as captured by the margin analysis in Sec.~\ref{sec:robustness}.

\subsection{Composability and small cascades}
\label{subsec:composability}

\begin{figure*}[t]
    \centering    \includegraphics[width=0.9\linewidth]{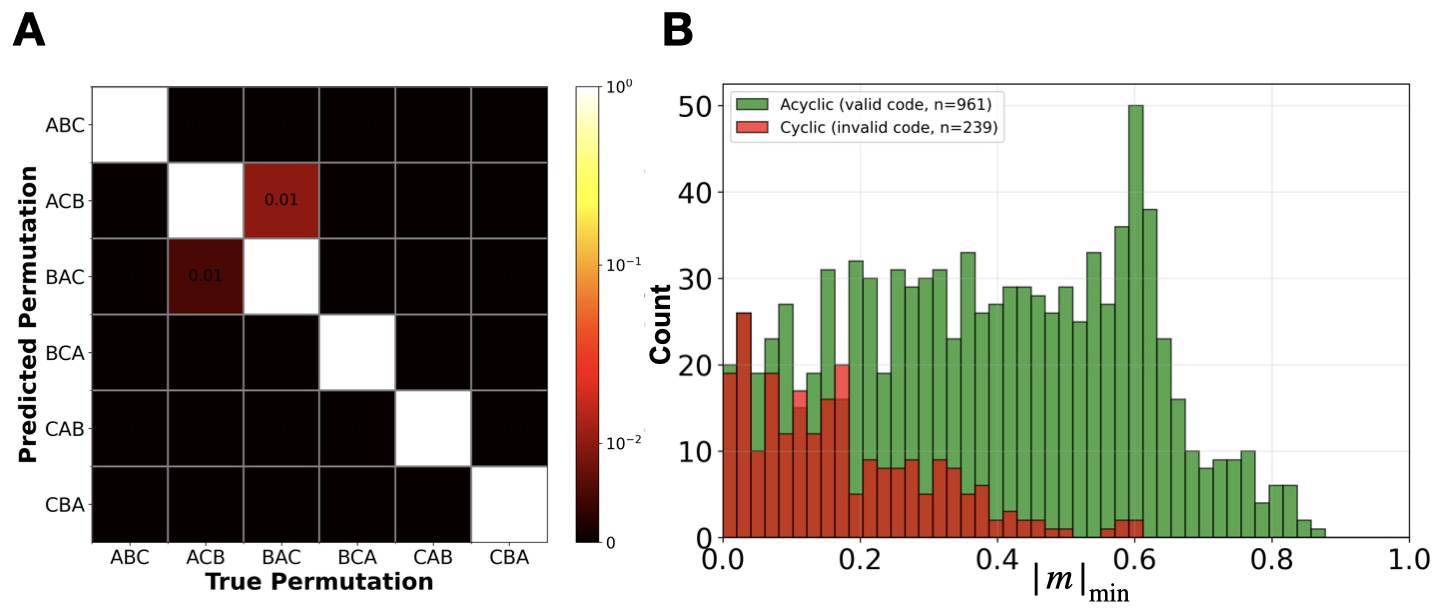}
 \caption{\textbf{Interferometric polychrony: three-spike permutation decoding via pairwise comparators and cycle repair.}
\textbf{(a)} Column-normalized confusion matrix for classifying the $3!=6$ temporal permutations of three input pulses $(A,B,C)$
using three pairwise temporal-order comparators (AB, BC, AC). Each comparator simulates the four-mode junction primitive and integrates
the readout intensities over a fixed window to obtain
$E_{L,R}=\int_{t_0}^{t_0+T_w}\! I_{L,R}(t)\,dt$ (with $t_0=\max(t_X,t_Y)+2\sigma$ and $T_w=6\sigma$),
returning a bit $b_{XY}=\mathbbm{1}[E_L>E_R]$ and signed margin
$m_{XY}=\log\!\bigl[(E_L+\epsilon)/(E_R+\epsilon)\bigr]$ with $\epsilon=10^{-12}$.
The three bits form a tournament code $(b_{AB},b_{BC},b_{AC})$; cyclic codes are repaired by flipping the weakest comparison,
i.e. the bit associated with $\arg\min\{|m_{AB}|,|m_{BC}|,|m_{AC}|\}$, after which the repaired code is decoded to a permutation. Color scale uses symmetric logarithmic normalization to reveal small off-diagonal values.
\textbf{(b)} Distribution of the minimum absolute margin
$|m|_{\min}=\min(|m_{AB}|,|m_{BC}|,|m_{AC}|)$ for acyclic (valid) and cyclic (invalid) codes before repair.
Cyclic codes concentrate at small $|m|_{\min}$, indicating that invalid codes predominantly occur near decision boundaries.
Unless noted otherwise, $200$ trials are generated per permutation (seed $42$) by jittering the two inter-spike gaps as
$g_{1,2}=\Delta t\,(1+0.10\,\mathcal{N}(0,1))$ with a minimum separation $\mathrm{min\_sep}=0.4$, forming ordered times
$(t_1,t_2,t_3)=(t_{\mathrm{ref}},t_{\mathrm{ref}}+g_1,t_{\mathrm{ref}}+g_1+g_2)$, and assigning them to $(t_A,t_B,t_C)$ according to the target permutation.
Comparator parameters are given in Supplement Sec.~S4.}
    \label{fig:N3_LUT}
\end{figure*}

We illustrate composition with the smallest nontrivial cascade: three spikes $(A,B,C)$ evaluated by three pairwise temporal-order primitives. Each primitive outputs a bit $b_{XY}$ and a signed margin $m_{XY}=\log(E_L/E_R)$ (Sec.~\ref{sec:circuits}), yielding a 3-bit tournament code $(b_{AB},b_{BC},b_{AC})$. In the noiseless limit the code is transitive and uniquely specifies one of the six strict orders. With jitter or static disorder, a marginal comparison can flip and create an intransitive 3-cycle, the minimal failure mode of composition. Table~\ref{tab:n3_tournament} lists the valid and cyclic codes; Fig.~\ref{fig:N3_LUT} quantifies the cycle rate and repair performance.

\begin{table}[t]
\centering
\begin{tabular}{c c c c l}
\toprule
Code $(b_{AB}\,b_{BC}\,b_{AC})$ & $AB$ & $BC$ & $AC$ & Order \\
\midrule
111 & $A\prec B$ & $B\prec C$ & $A\prec C$ & $ABC$ \\
101 & $A\prec B$ & $C\prec B$ & $A\prec C$ & $ACB$ \\
011 & $B\prec A$ & $B\prec C$ & $A\prec C$ & $BAC$ \\
010 & $B\prec A$ & $B\prec C$ & $C\prec A$ & $BCA$ \\
100 & $A\prec B$ & $C\prec B$ & $C\prec A$ & $CAB$ \\
000 & $B\prec A$ & $C\prec B$ & $C\prec A$ & $CBA$ \\
\addlinespace[0.6em]
110 & $A\prec B$ & $B\prec C$ & $C\prec A$ & cycle \\
001 & $B\prec A$ & $C\prec B$ & $A\prec C$ & cycle \\
\bottomrule
\end{tabular}
\caption{\textbf{Tournament codes for $N=3$.}
Bits are $b_{XY}=\mathbbm{1}[m_{XY}>0]$ with margin $m_{XY}=\log(E_L/E_R)$, where the primitive routes left when $t_X<t_Y$.
The six acyclic codes correspond to strict total orders; $110$ and $001$ are intransitive 3-cycles.}
\label{tab:n3_tournament}
\end{table}

As in LUT-style temporal hashing \cite{Izhikevich2025SpikingManifesto}, many comparisons can be performed in parallel and their bits concatenated into a higher-entropy address. Composition must respect the wrap-free and coherence constraints of Sec.~\ref{sec:phase_delay}; violations introduce aliasing or contrast loss that cannot be corrected downstream.

\paragraph{Three-spike tournament codes.}
For $(A,B,C)$, the three comparators produce
$
b_{XY}=\mathbbm{1}[m_{XY}>0],
$
with the convention that routing left corresponds to $t_X<t_Y$. The six acyclic codes correspond to strict total orders, while $110$ and $001$ are intransitive cycles (Table~\ref{tab:n3_tournament}).

\paragraph{Cycle repair from margins.}
Cycles arise from near ties. A local repair restores transitivity by flipping the least confident comparison,
\begin{equation}
\text{flip}\;\; \arg\min\{|m_{AB}|,\;|m_{BC}|,\;|m_{AC}|\}.
\end{equation}
Let $|m|_{\min}=\min\{|m_{AB}|,|m_{BC}|,|m_{AC}|\}$. Cyclic outcomes therefore concentrate at small $|m|_{\min}$, while away from decision boundaries the code is transitive and decodes correctly without repair. Figure~\ref{fig:N3_LUT} confirms both trends.

Importantly, resolving the least stable comparison does not require explicit digital processing. The margins $m_{XY}$ are analog observables available at readout, and small $|m_{XY}|$ naturally corresponds to slower saturation or higher susceptibility to competition. Local gain biasing, finite integration windows, or weak coupling among comparator outputs can therefore preferentially suppress or flip the weakest comparison, relaxing the system toward a transitive code. Explicit post-readout margin comparison is a convenient option, not a requirement.

This provides a composability rule: each primitive outputs a discrete bit together with a confidence margin that flags near-tie failures and supports local repair before final digitization. Physically, the primitive already performs routing by selecting port $L$ or $R$ (Fig.~\ref{fig:primitives}); fan-out is implemented by splitting the winning channel.

\subsection{How the primitives map onto LUT-style computation}
\label{subsec:lut_mapping}

The circuits above realize the address-evaluation step in LUT-style spiking computation \cite{Izhikevich2025SpikingManifesto}:
\begin{itemize}
\item \emph{COMPARE:} temporal comparators output a one-bit order/hash of relative delay and a signed confidence margin;
\item \emph{SCORE:} interferometric lookup (Sec.~\ref{sec:linear_lookup}) evaluates $K$ candidate templates in parallel by phase correlation;
\item \emph{SELECT:} the driven--dissipative WTA stage (Sec.~\ref{sec:wta}) implements a physical $\arg\max$, producing a one-hot address and routing the winner.
\end{itemize}
This pipeline performs temporal hashing and hard top-1 selection directly in the wave domain, without per-spike timestamping or digital delay subtraction.

Our goal is narrower than general computation: a reusable, composable set of timing-native primitives that outputs discrete addresses for downstream lookup. Composition is limited by the operating envelope of Sec.~\ref{sec:phase_delay} (wrap-free window, phase coherence across the span, and reliable competitive readout). Scaling beyond small cascades therefore depends on (i) preserving separability under cross-talk and template crowding, (ii) maintaining programmability via calibration against static offsets and drift, and (iii) defining inter-stage interfaces (re-encoding a selected index for subsequent stages and implementing value retrieval). Within these constraints, phase-coded temporal indexing offers a direct physical route to fast address selection.

\section{Robustness, noise, scaling and hardware-in-the-loop optimization}
\label{sec:robustness}

For an $\arg\max$ (WTA) decision, the patterns most sensitive to perturbations are those with a small top-two gap.
We quantify two ingredients that control this sensitivity: (i) phase noise, which reduces the coherent part of the matched score, and (ii) statistical ``crowding'' from the template library, which sets the scale of the strongest incorrect response.
Throughout we use the amplitude margin
$\Delta_{\rm lin}(x)=A_{k^\ast}(x)-\max_{k\neq k^\ast}A_k(x)$
[Eq.~\eqref{eq:Delta_lin_def}] with $A_k\equiv|\Psi_k|$.

\subsection{Phase noise from jitter, static disorder, and finite coherence}
\label{subsec:jitter_coherence}

In the rotating-frame encoding, a timing error $\delta t_j$ on input $j$ produces a phase error
$\delta\phi_j=\Omega\,\delta t_j$ (the sign is immaterial for the statistics below).
We separate three contributions to the per-term phase error entering the linear phasor sum:
(i) timing jitter with rms $\sigma_t$ [Eq.~\ref{eq:jitter_phase}].
(ii) static phase offsets in the programmed couplings,
\begin{equation}
\theta_{jk}\mapsto \theta_{jk}+\delta\theta_{jk},\quad
\mathbb{E}[\delta\theta_{jk}]=0,\;\;\mathrm{Var}(\delta\theta_{jk})=\sigma_\theta^2,
\end{equation}
where $\delta\theta_{jk}$ is fixed within a given device realization; and
(iii) finite-coherence dephasing (e.g.\ phase diffusion) accumulated over the decision window, with variance $\sigma_{\rm coh}^2$ \cite{Carusotto2013RMP}.
For analytic estimates we lump these into a single additive phase error per term and assume the three mechanisms are independent, so the variances add:
\begin{equation}
\sigma_{\rm eff}^2 \equiv \sigma_\phi^2 + \sigma_\theta^2 + \sigma_{\rm coh}^2 .
\label{eq:sigma_eff_def}
\end{equation}
In practice, $\sigma_\theta$ should be interpreted as the residual static phase error after any calibration (see below).

\paragraph{Matched-channel coherence under phase noise.}
Fix the matched index $k^\ast$ and choose the reference phases so that, in the absence of errors, all $N$ contributions add in phase.
Write the matched score as
\begin{equation}
\Psi_{k^\ast}=\sum_{j=1}^{N} a_j e^{i\varepsilon_j},
\qquad a_j\equiv |J_{jk^\ast}|,
\end{equation}
where $\varepsilon_j$ is the net phase error on term $j$.
Assume $\varepsilon_j$ are independent, zero-mean Gaussians with variance $\sigma_{\rm eff}^2$ \footnote{For dynamic noise (jitter and coherence), the expectation is over repeated trials. For static offsets, the same expressions describe the typical suppression when averaging over an ensemble of random $\delta\theta_{jk^\ast}$, or equivalently when treating the residual static mismatch across $j$ as effectively random.}.
Then $\mathbb{E}[e^{i\varepsilon_j}]=e^{-\sigma_{\rm eff}^2/2}$ and standard phasor statistics \cite{Goodman2015StatisticalOptics} give
\begin{align}
\mathbb{E}[\Psi_{k^\ast}]
&=
e^{-\sigma_{\rm eff}^2/2}\sum_{j=1}^N a_j,
\label{eq:mean_Psi_matched}\\[2pt]
\mathbb{E}[|\Psi_{k^\ast}|^2]
&=
(1-e^{-\sigma_{\rm eff}^2})\sum_{j=1}^N a_j^2
+
e^{-\sigma_{\rm eff}^2}\Bigl(\sum_{j=1}^N a_j\Bigr)^2 .
\label{eq:power_Psi_matched}
\end{align}
Equation~\eqref{eq:power_Psi_matched} makes the separation explicit: a coherent component (the squared sum) suppressed by $e^{-\sigma_{\rm eff}^2}$ and an incoherent floor (the sum of squares) that remains even when the coherent term is lost.

\paragraph{Static disorder and calibration.}
Unlike jitter and dephasing, $\delta\theta_{jk}$ is time-independent within a device.
If the hardware supports phase programmability, these offsets can be estimated and compensated by closed-loop tuning that maximizes measured margins (cf.\ self-configuring interferometers \cite{Miller2013SelfConfiguring}).
After such calibration, $\sigma_\theta$ in Eq.~\eqref{eq:sigma_eff_def} should be understood as the residual mismatch set by drift, cross-talk, and measurement noise.

\subsection{Margin-controlled error probability}
\label{subsec:margin_error}

Let $A_k=|\Psi_k|$ and denote the winner and runner-up by $A_{k^\ast}$ and $A_{\rm ru}\equiv\max_{k\neq k^\ast}A_k$, so that
$\Delta_{\rm lin}=A_{k^\ast}-A_{\rm ru}$.
Exact misaddressing probabilities depend on the joint fluctuations of $A_{k^\ast}$ and $A_{\rm ru}$.
As a gap-to-noise proxy, model the perturbation of the winner--runner-up gap by a zero-mean random variable $\delta\Delta$ with rms $\sigma_\Delta$ and define
\begin{equation}
P_{\rm error}\equiv \Pr\!\left[\Delta_{\rm lin}+\delta\Delta<0\right].
\label{eq:Perror_def}
\end{equation}
If $\delta\Delta\sim\mathcal{N}(0,\sigma_\Delta^2)$, then
\begin{equation}
P_{\mathrm{error}}
= \frac{1}{2}\,\mathrm{erfc}\!\left(\frac{\Delta_{\mathrm{lin}}}{\sqrt{2}\,\sigma_\Delta}\right)
\le \frac{1}{2}\exp\!\left(-\frac{\Delta_{\mathrm{lin}}^{2}}{2\sigma_\Delta^{2}}\right),
\label{eq:margin_error_bound}
\end{equation}
where the inequality is a loose but monotone tail bound for $\Delta_{\mathrm{lin}}>0$.
We use Eq.~\eqref{eq:margin_error_bound} only as an interpretable margin-to-noise proxy; all reported accuracies are obtained from the driven--dissipative simulations (Secs.~\ref{sec:wta}, \ref{sec:circuits} and Supplemental Material).

\subsection{Scaling with input dimension $N$ and number of templates $K$}
\label{subsec:scaling}

This subsection gives reference scaling in a fully dephased (random-phase) competitor model, and then summarizes correlated template libraries through an effective competitor count $K_{\rm eff}$ (Supplement Sec.~S2). \paragraph{Spectral effective competitor count.}
To quantify template diversity (and hence the number of effective competitors), we use a spectral effective count
$K_{\rm eff}^{\rm spec}$ computed from the normalized template overlap matrix.
Specifically, define normalized template phasor vectors $v^{(k)}\in\mathbb{C}^N$ by
$v^{(k)}_j \equiv N^{-1/2}e^{-i\Omega t^{(k)}_j}$ and form the nonnegative overlap matrix
$$
G_{kk'} \equiv \left|\langle v^{(k)},v^{(k')}\rangle\right|^2
= \left|\frac{1}{N}\sum_{j=1}^N e^{i\Omega\left(t^{(k)}_j-t^{(k')}_j\right)}\right|^2 ,
$$
which is symmetric and positive semidefinite, hence has eigenvalues $\lambda_\alpha\ge 0$.
We then form the normalized eigenvalues $p_\alpha=\lambda_\alpha/\sum_\beta \lambda_\beta$ and define
\begin{equation}
K_{\rm eff}^{\rm spec}\equiv \frac{1}{\sum_\alpha p_\alpha^2}.
\label{eq:keff_spec_main}
\end{equation}
This participation-ratio measure satisfies $1\le K_{\rm eff}^{\rm spec}\le K$, approaching $K$ for diverse (weakly correlated) libraries
and $\approx 1$ for maximally crowded libraries.

\paragraph{Matched channel.}
For near-uniform magnitudes $|J_{jk^\ast}|\approx a$,
Eq.~\eqref{eq:power_Psi_matched} implies the rms matched amplitude
\begin{eqnarray}
A_{k^\ast}&\sim &a\,\sqrt{N+(N^2-N)e^{-\sigma_{\rm eff}^2}}\nonumber \\
&\approx&
\begin{cases}
a\,N\,e^{-\sigma_{\rm eff}^2/2}, & N e^{-\sigma_{\rm eff}^2}\gg 1,\\[2pt]
a\,\sqrt{N}, & e^{-\sigma_{\rm eff}^2}\ll 1,
\end{cases}
\label{eq:Ak_matched_scaling}
\end{eqnarray}
where the first line is the coherence-dominated regime and the second is the fully dephased limit.

\paragraph{Strongest competitor: dephased reference.}
In the random-phase limit, an unrelated template produces a sum of $N$ phasors with approximately zero mean and variance $\sim a^2N$, so $A_k$ is Rayleigh-like with a tail $\Pr[A_k>t]\approx \exp(-t^2/(a^2N))$.
For $K-1$ approximately independent competitors, a union bound gives
\begin{equation}
\Pr\!\left[\max_{k\neq k^\ast}A_k>t\right]\;\lesssim\; (K-1)\exp\!\left(-\frac{t^2}{a^2N}\right),
\end{equation}
so the runner-up scale is
\begin{equation}
\max_{k\neq k^\ast} A_k \sim a\,\sqrt{N\log K},
\label{eq:max_spurious_scaling}
\end{equation}
up to order-unity constants \cite{Vershynin2018HighDimProb} (see Appendix~\ref{app:analytics}). Eq.~(\ref{eq:max_spurious_scaling}) is only a baseline; real template libraries often violate the i.i.d. phase assumption.

\paragraph{Effective crowding and margin estimate.}
At low dephasing the competitor channels are generally neither independent nor fully random-phase; near-duplicate templates can also yield partially coherent responses.
We summarize these effects by defining $K_{\rm eff}$ operationally from the observed runner-up statistics (Supplement Sec.~S2) and replacing $K\mapsto K_{\rm eff}$ in Eq.~\eqref{eq:max_spurious_scaling}.
This motivates the typical-margin estimate
\begin{equation}
\Delta_{\rm lin}\sim
a\left( N e^{-\sigma_{\rm eff}^2/2} - c\,\sqrt{N\log K_{\rm eff}(N)} \right),
\label{eq:margin_scaling}
\end{equation}
with $c=\mathcal{O}(1)$ absorbing the extreme-value prefactor.
Equation~\eqref{eq:margin_scaling} is meaningful only when the matched channel remains coherence-dominated; otherwise the first term crosses over to $\sim a\sqrt{N}$ [Eq.~\eqref{eq:Ak_matched_scaling}] and discrimination is lost for large enough $K_{\rm eff}$.
WTA does not alter these interferometric scalings; it only maps a finite set of analog amplitudes $\{A_k\}$ to a discrete address.


In summary, phase noise enters the linear score through the total phase variance $\sigma_{\rm eff}^2$ [Eq.~\eqref{eq:sigma_eff_def}], suppressing coherent contrast in the matched channel as in Eqs.~\eqref{eq:mean_Psi_matched}--\eqref{eq:power_Psi_matched}.
In a fully dephased reference model, the runner-up grows as $\sqrt{N\log K}$ [Eq.~\eqref{eq:max_spurious_scaling}].
Correlated libraries are captured by an empirical $K_{\rm eff}$, leading to the margin estimate in Eq.~\eqref{eq:margin_scaling}.
Static phase offsets are time-independent and therefore, in principle, reducible by calibration; $\sigma_\theta$ should be interpreted as the residual after such tuning \cite{Miller2013SelfConfiguring}.

\section{Compilation and hardware-in-the-loop calibration}
\label{sec:calibration}

The interferometric scorer is programmable, but its effective phases and gains are not identical across nominally identical devices. Fabrication offsets, drift, and nonuniform pumping shift the score landscape and can bias a discrete address map if left uncorrected. In coherent interferometer hardware, such closed-loop tuning is routine \cite{Miller2013SelfConfiguring,Clements2016Optica,PerezLopez2020SelfConfigurationPICs,xu2022self}. Here we use the same idea, but with a calibration target matched to our task: reliable top-1 selection. Concretely, we (i) compile a desired temporal template library into nominal phase settings, and (ii) refine those settings on the realized device by maximizing an output-level margin objective using intensity-only readout. The procedure treats the device as a black box; it does not require phase-sensitive measurements or explicit identification of internal errors.

\subsection{Forward compilation: temporal templates to phase programs}
\label{subsec:forward_compile}

Given a template bank $\{t_j^{(k)}\}_{j=1}^N$ for each address $k$, the ideal mapping sets the programmed phases to
\begin{equation}
\theta_{jk}^{(0)}=\Omega\, t_j^{(k)},
\label{eq:compile_ideal}
\end{equation}
so that the intended template is aligned in the interferometric score $\Psi_k$ (Sec.~\ref{sec:linear_lookup}). On hardware, the programmed parameters $\{\theta_{jk},|J_{jk}|\}$ are realized with device-specific offsets. A minimal static model is
\begin{equation}
\theta_{jk}^{\mathrm{eff}}=\theta_{jk}+\delta\theta_{jk},\qquad
|J_{jk}^{\mathrm{eff}}|=|J_{jk}|(1+\delta a_{jk}),
\label{eq:static_disorder_model}
\end{equation}
where $\delta\theta_{jk}$ and $\delta a_{jk}$ are fixed within a device instance (and may drift slowly in time). The role of calibration is to choose controls $\{\theta_{jk}\}$ that recover the desired addressing behavior under $\theta^{\mathrm{eff}},J^{\mathrm{eff}}$, without first estimating $\delta\theta_{jk}$ or $\delta a_{jk}$.

\subsection{Calibration objective: maximize a measurable post-readout margin}
\label{subsec:margin_objective}

Calibration is defined on quantities available at the outputs of the selector. Let $I_k$ denote a robust per-port observable (e.g., steady-state intensity or integrated output energy over a fixed window). For a labeled calibration input whose correct address is $k^\ast$, we use the intensity margin $M$ [Eq.~\ref{eq:M_def}] and define the loss
\begin{equation}
\mathcal{L}(\{\theta_{jk}\})=-M.
\label{eq:margin_loss}
\end{equation}
This choice matches the failure mode of discrete selection: errors occur when the winner and nearest competitor are close. Increasing $M$ enlarges the separation at the measurement plane, which improves tolerance to residual static mismatch and to dynamic perturbations that act through the same near-tie mechanism.

\paragraph{A smooth surrogate (optional).}
When a differentiable approximation is convenient (e.g., for gradient-based updates with noisy readout), the runner-up can be softened by replacing the non-smooth maximum with a log-sum-exp relaxation,
\begin{equation}
\max_{k\neq k^\ast} I_k
\;\approx\;
\frac{1}{\beta}\log\!\left(\sum_{k\neq k^\ast} e^{\beta I_k}\right),
\label{eq:logsumexp}
\end{equation}
with $\beta$ setting the trade-off between smoothness and fidelity to the true runner-up.

\begin{figure*}[t]
  \centering  \includegraphics[width=0.95\textwidth]{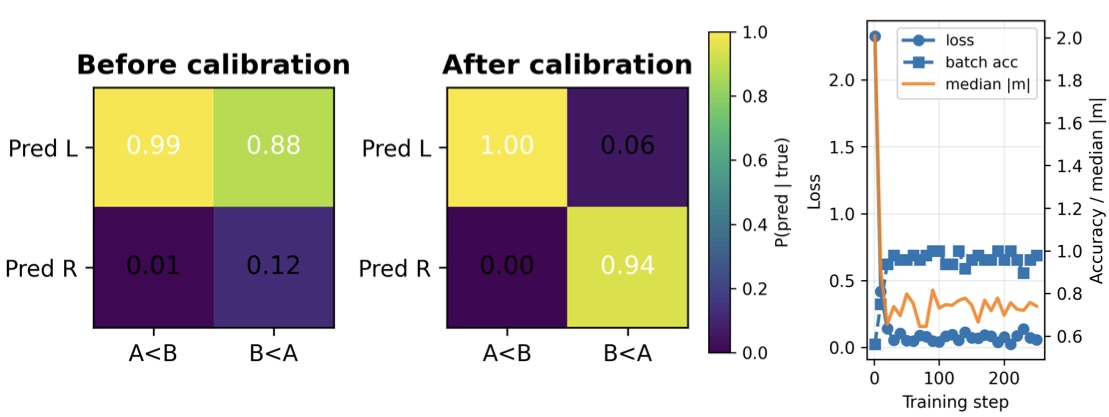}
\caption{\textbf{Hardware-in-the-loop calibration rescues the $N=2$ temporal-order primitive under fabrication mismatch.}
A four-mode temporal comparator (modes $A,B$ inputs; $L,R$ outputs) is degraded by strong static fabrication disorder:
unknown coupling-phase offsets $\theta_{\mathrm{fab}}\sim\mathcal{N}(0,\sigma_{\mathrm{fab}}^2)$ with $\sigma_{\mathrm{fab}}=1.5~\mathrm{rad}$ (the plotted device is one fixed random draw from this distribution, so its empirical RMS need not equal $\sigma_{\mathrm{fab}}$), amplitude imbalance via lognormal coupling scales (25\% RMS, clipped to $[0.5,1.5]$),
and detuning scatter $\delta\Delta,\delta\omega_0\sim\mathcal{N}(0,(0.15J)^2)$ applied to $(\Delta,\omega_0)$. 
\textbf{Left:} column-normalized confusion matrix $P(\mathrm{pred}\mid \mathrm{true})$ on a held-out test set of $800$ trials
($400$ per class), before calibration with $\boldsymbol{\theta}_{\mathrm{ctrl}}=\mathbf{0}$ (mean diagonal accuracy $55.9\%$; failure mode:
nearly always predicts LEFT).
\textbf{Middle:} after $250$ SPSA steps (batch size $48$; constant learning rate $\xi=0.08$ and perturbation $\delta=0.08~\mathrm{rad}$;
$\ell_2$ regularization $\lambda=10^{-4}$; softmax temperature $\beta=1$), phase-only programming learns control phases
$\boldsymbol{\theta}_{\mathrm{ctrl}}=[-0.065,\,-1.236,\,-0.012,\,+0.802]~\mathrm{rad}$ and restores routing
(mean diagonal accuracy $97.2\%$).
\textbf{Right:} training dynamics (loss, batch accuracy, and median $|m_{AB}|$ versus step), showing monotonic loss decrease and rapid accuracy recovery.
Training examples use $\pm30\%$ variation in the nominal separation $\delta t=1.60/J$ with a minimum separation $0.8\sigma$ and timing jitter
$\sigma_t=0.15\sigma$ (pulse width $\sigma=0.25/J$); the device is simulated in the linear regime ($U=0$) with
$J=1$, $\gamma=0.1J$, $\Delta_0=\omega_{0,0}=1.2J$, and $\kappa=0.5J$.
Calibration uses only black-box integrated output scores and does not assume a device model; full disorder and training details are given in Supplement Sec.~S5.}
  \label{fig:calibration}
\end{figure*}

\paragraph{Why the objective is margin-based.}
Because the address map is discontinuous at decision boundaries, a loss on the hard index provides little gradient information away from misclassified points. The margin objective instead pushes boundary-adjacent queries away from their nearest competitor, directly targeting stability of the selected address under noise and residual disorder. Figure~\ref{fig:calibration} illustrates this on the two-spike comparator (Sec.~\ref{subsec:order_primitive}): static mismatch can create a biased left/right map, and intensity-only phase tuning restores correct routing by reopening the measured margin distribution.

\subsection{Hardware-in-the-loop optimization}
\label{subsec:calibration}

We implement calibration as a hardware-in-the-loop (HIL) optimization: apply a candidate phase program, present calibration patterns, measure $\{I_k\}$, update the phases, and repeat
\cite{Comsa2022TemporalCodingTNNLS,Sun2023LearnableAxonalDelay,Hammouamri2024DelayLearningICLR}.
Related in-situ training loops have been demonstrated in wave processors \cite{Onodera2025MultimodeWaveTraining,Hughes2018Optica,pai2023experimentally}; our difference is the objective (robust discrete selection rather than regression fidelity) and the restriction to intensity-only feedback.

\paragraph{Finite-difference updates.}
For a single control parameter $\theta_p$ (one programmable phase), a central-difference estimate is
\begin{equation}
\frac{\partial \mathcal{L}}{\partial \theta_p}
\;\approx\;
\frac{\mathcal{L}(\theta_p+\delta)-\mathcal{L}(\theta_p-\delta)}{2\delta}.
\label{eq:fd_grad}
\end{equation}
With $P\sim NK$ phase parameters, one full gradient step costs $\mathcal{O}(P)$ loss evaluations. In practice, repeated measurements (or mini-batches of inputs) can be used to average readout noise.

\paragraph{Simultaneous perturbation (SPSA).}
When $P$ is large, SPSA estimates the full gradient using two loss evaluations per iteration \cite{Spall1992TAC}. Let $\theta\in\mathbb{R}^P$ collect all phases and let $\boldsymbol{\xi}\in\{-1,+1\}^P$ be an i.i.d.\ Rademacher vector. The $p$-th component of the gradient estimate is
\begin{equation}
\bigl[\widehat{\nabla \mathcal{L}}(\theta)\bigr]_p
=
\frac{\mathcal{L}(\theta + d\,\boldsymbol{\xi}) - \mathcal{L}(\theta - d\,\boldsymbol{\xi})}{2d\,\xi_p}\,,
\label{eq:spsa}
\end{equation}
followed by the update $\theta\leftarrow \theta-\alpha\,\widehat{\nabla \mathcal{L}}$ with step size $\alpha>0$.

\paragraph{What the loop is actually correcting.}
The loop compensates static device-specific distortions of the score landscape (e.g., fixed phase offsets, coupling imbalance, and slowly varying drift) by reprogramming phases so that the measured post-readout margins match the intended behavior. It does not require reconstructing $\delta\theta_{jk}$ or $\delta a_{jk}$, and it does not assume access to internal fields.

\subsection{Scope and platform considerations}
\label{subsec:calibration_limits}

The requirements are minimal: (i) phase-like controls that affect the interferometric scorer, and (ii) per-port intensity readout. The procedure addresses static mismatch, but it does not remove trial-to-trial timing jitter or finite-coherence phase diffusion, which remain as dynamic noise sources. Moreover, patterns that are intrinsically near a decision boundary have small margins by definition; no static reprogramming can make those cases error-free for all noise realizations. Calibration is therefore statistical: it reshapes the margin distribution on a representative calibration set to reduce the fraction of near-tie queries and to lower the measured misaddressing rate under the target noise budget.

In summary, the combination of forward compilation [Eq.~\eqref{eq:compile_ideal}] and HIL margin optimization [Eq.~\eqref{eq:margin_loss}] provides a reproducible way to program the timing-to-index primitive across devices using intensity-only feedback, with measurable success criteria (accuracy and margin distributions after calibration).

\section{Platforms and systems role}
\label{sec:platforms_outlook}

PWC relies on three ingredients: phase encoding of spike times in a rotating frame (Sec.~\ref{sec:phase_delay}),
parallel scoring by coherent interference (Sec.~\ref{sec:linear_lookup}), and digitization to a single address by nonlinear competition (Sec.~\ref{sec:wta}).
Here we summarize the physical requirements, list candidate platforms, and relate device-level routing errors to task-level impact using a simple benchmark.

\subsection{Physical requirements}
\label{subsec:platform_requirements}

The operating envelope is set by four requirements.

\paragraph{Phase reference and wrap-free window.}
Encoding uses $\phi=\Omega t \bmod 2\pi$ over a decision window, requiring
$ t_{\max} < T_{\mathrm{wrap}}=2\pi/\Omega$ (Sec.~\ref{sec:phase_delay}).
$\Omega$ is set by an external reference (detuning, modulation tone, injection locking) rather than by fine resonance.
This choice trades delay span against sensitivity: timing jitter maps to phase noise $\sigma_\phi=\Omega\sigma_t$ (Sec.~\ref{sec:robustness}).

\paragraph{Coherent linear superposition with programmable complex weights.}
Scoring is linear and phase sensitive,
so phase must remain meaningful across the delay span, i.e., $\tau_{\mathrm{coh}}\gtrsim  t_{\max}$ (Sec.~\ref{sec:phase_delay}).
Finite coherence suppresses interferometric contrast continuously and reduces margins (Sec.~\ref{sec:robustness}).
Implementation therefore requires tunable complex couplings (or an equivalent phase-programming mechanism) and calibration of static phase/amplitude errors, as in coherent photonic processors \cite{Miller2013SelfConfiguring,Clements2016Optica}.

\paragraph{Nonlinear WTA digitization.}
Digitization requires an effective $\arg\max$ implemented by competitive gain/loss or shared depletion, selecting a single mode $k^\ast$ (Sec.~\ref{sec:wta}).
Such selection is common in gain-dissipative oscillator networks and polariton graphs \cite{Berloff2017NatMater,Kalinin2019PolaritonicNetwork}.

\paragraph{Reset and readout.}
Between inferences the system must return reproducibly to baseline (passive decay or active quench) and provide an observable that identifies the winner (emitted intensity or time-integrated energy).

\subsection{Candidate physical platforms}
\label{subsec:candidate_platforms}

We now map the requirements in Sec.~\ref{subsec:platform_requirements} onto concrete hardware. Below we highlight two implementation routes that naturally supply these ingredients, and we summarize several additional candidate substrates in Appendix~\ref{app:platforms}.

\paragraph{Exciton--polariton networks.}
Exciton--polaritons offer interferometric access to the complex field, low-threshold nonlinearities, and driven--dissipative condensation \cite{Carusotto2013RMP,Wei2022RoomTempOrganicPolaritons}.
Programmable ``polariton graphs'' demonstrate controllable coupling and phase relations in driven oscillator networks \cite{Berloff2017NatMater,Kalinin2019PolaritonicNetwork}, and spiking-like temporal dynamics (including leaky integrate--and--fire behavior) have been reported in polariton microstructures \cite{Tyszka2023LIFPolaritons}.
Reservoir-mediated delayed nonlinearities have also been used to realize ultrafast time-domain primitives, including picosecond-scale time-coded processing \cite{Mirek2022TimeDelayedNN,Mirek2021NanoLett}.

For timing budgets, two intrinsic scales matter: the polariton lifetime and the reservoir lifetime.
Representative devices report polariton decay times above threshold on the order of $\sim 10$~ps and reservoir lifetimes on the order of $\sim 100$~ps \cite{Tyszka2023LIFPolaritons}, placing the native dynamics in the sub-nanosecond regime (device-dependent).
Energy figures in the literature are typically quoted per excitation or per spike; translating them to per-decision energy for an $N$-input, $K$-way selector requires accounting for fan-in, readout, and reset.

\paragraph{Programmable photonic scoring with laser-based WTA.}
A practical split is to implement the calibrated linear scorer $J_{jk}$ in a programmable photonic substrate (e.g., interferometric meshes or reconfigurable multimode/slab processors) and implement WTA digitization in a fast nonlinear oscillator or laser bank.
Large coherent linear transforms have been demonstrated in integrated photonics \cite{Shen2017NatPhoton,Wetzstein2020Nature}, and multimode/slab processors provide dense mixing without explicit mesh architectures \cite{Onodera2025MultimodeWaveTraining}.
Many such platforms still extract a discrete decision downstream (electronically or algorithmically); here we emphasize pairing wave-level scoring with a native driven--dissipative competitive stage so that the module emits a one-hot decision.

VCSEL-based photonic neurons and VCSEL--SA networks provide one candidate route to the nonlinear competitive stage \cite{Skalli2022OMEVCSEL,Zhang2020JLTWTA}.
Reported bandwidth and efficiency metrics in the VCSEL literature are promising, but mapping them to energy and latency per discrete WTA decision depends on biasing, coupling losses, and the required readout margin.

\paragraph{Other candidate substrates.}
The same ingredients (phase reference, programmable interference, and competitive digitization) may also be realized in other wave and oscillator media, including coherent Ising-machine style oscillator networks, magnonic spin-wave processors, RF/microwave oscillator networks, acoustic/phononic metamaterials with active gain, and superconducting microwave circuits.
We summarize these options and their key trade-offs in Appendix~\ref{app:platforms}. A minimal proof-of-principle is a $K=4$ address selector that demonstrates (i) phase-coded scoring from latency, and (ii) WTA digitization with a measured margin distribution under injected timing jitter.

\subsection{Engineering challenges and mitigation strategies}
\label{subsec:eng_challenges}

\paragraph{Timing precision and reference distribution.}
Because $\sigma_\phi=\Omega\sigma_t$, the timing budget (pulse generation, delivery, detection) sets achievable margins.
Co-design of $\Omega$, $ t_{\max}$, and $N$ trades delay span against robustness at fixed error rate (Sec.~\ref{sec:robustness}).

\paragraph{Separating linear scoring from nonlinear selection.}
During scoring, nonlinearities should not distort $\Psi_k$ (e.g., saturation or history dependence); otherwise the effective template map drifts and becomes difficult to calibrate.
This can be enforced by pump scheduling, staged coupling, or explicit score versus select windows (Secs.~\ref{sec:linear_lookup}--\ref{sec:wta}).

\paragraph{Static disorder, drift, and programmability.}
Static phase offsets and coupling inhomogeneities shift decision boundaries and reduce margins.
They can be addressed by compiling to the measured hardware and closing the loop with hardware-in-the-loop calibration (Sec.~\ref{sec:calibration}), as in programmable coherent photonics \cite{Miller2013SelfConfiguring,Clements2016Optica}.

\paragraph{I/O and system bottlenecks.}
Even with fast indexing, throughput can be limited by how values are stored and moved once an address is chosen.
A near-term systems role is a selection/routing coprocessor that outputs a discrete index (or one-hot port) to downstream memory or neuromorphic blocks.
A platform-specific energy/latency model comparing digitize--compare--select pipelines with interference+WTA indexing, including triggered memory access, is needed to identify break-even regimes in $(N,K)$ and timing/readout budgets.

\subsection{Systems role: relation to attention and routing}
\label{subsec:attention_routing}

Sparse attention and mixture-of-experts (MoE) architectures implement discrete routing: selecting a top-1 (or top-$k$) expert index for each query \cite{vaswani2017attention,shazeer2017outrageously}.
In spike-timing formalisms, the analogous operation is an address map from a spatiotemporal spike pattern to a discrete index \cite{Izhikevich2006Polychronization,Izhikevich2025SpikingManifesto}.
The practical bottleneck for ultrafast physical substrates is computing this index at native timescales without converting events into timestamps, bins, or rates.

\paragraph{Routing-only MoE benchmark.}
We quantify system-level sensitivity to misrouting using a minimal hard top-1 MoE gate in which expert computation is fixed and only the routed index is corrupted.
For each dataset seed, sample cluster means $\{\mu_k\}_{k=1}^K\subset\mathbb{R}^{10}$ and hold them fixed.
Each test sample draws a ground-truth expert label $k^\ast\sim\mathrm{Unif}\{1,\ldots,K\}$ and input
\begin{equation}
x \sim \mathcal{N}(\mu_{k^\ast},\sigma_x^2 I),
\label{eq:moe_x_gen_main}
\end{equation}
and receives a binary label from an expert-specific linear rule
\begin{equation}
y=\mathbbm{1}\!\left[\mathbf{w}_{k^\ast}^\mathsf{T}x+b_{k^\ast}>0\right],
\label{eq:moe_y_rule_main}
\end{equation}
with $\{(\mathbf{w}_k,b_k)\}_{k=1}^K$ sampled once per seed.
Expert $k$ is a logistic-regression classifier trained only on samples with label $k^\ast=k$, enforcing genuine specialization.

\paragraph{Injected routing error and device mapping.}
Routing is modeled as a stochastic top-1 selector with controllable misrouting probability
$
p_{\mathrm{route}}\equiv \Pr(\hat{k}\neq k^\ast),$ where
$\hat{k}=k^\ast$ with probability $1-p_{\mathrm{route}}$ and $\hat{k}=
\text{Uniform}\!\big(\{1,\ldots,K\}\setminus\{k^\ast\}\big),$ with probability $ p_{\mathrm{route}}$.
Uniform misrouting is a deliberately simple baseline. To match measured device behavior more closely,
one can instead use an empirical confusion matrix
Let
\begin{equation}
C_{k' \mid k}
\;\equiv\;
\Pr\!\left( \hat{k} = k' \,\middle|\, k^\ast = k \right),
\end{equation}
denote the empirically measured routing confusion matrix obtained from repeated trials.
Given a true label $k^\ast$, the observed output is sampled as
$
\hat{k} \sim C_{\cdot \mid k^\ast}.
$
The scalar error model discussed above corresponds to
\begin{equation}
C_{k^\ast \mid k^\ast} = 1 - p_{\mathrm{route}},
\qquad
C_{k' \mid k^\ast} = \frac{p_{\mathrm{route}}}{K-1},
\quad k' \neq k^\ast .
\end{equation}

Given $\hat{k}$, the system prediction is the output of expert $\hat{k}$, so the end-to-end accuracy is
\begin{equation}
P_{\mathrm{task}}(p_{\mathrm{route}})\equiv \Pr\!\left(\widehat{y}(x)=y\right),
\label{eq:moe_task_acc_main}
\end{equation}
evaluated as a function of $p_{\mathrm{route}}$ with all experts held fixed.

\begin{figure*}[t]
\centering
\includegraphics[width=\textwidth]{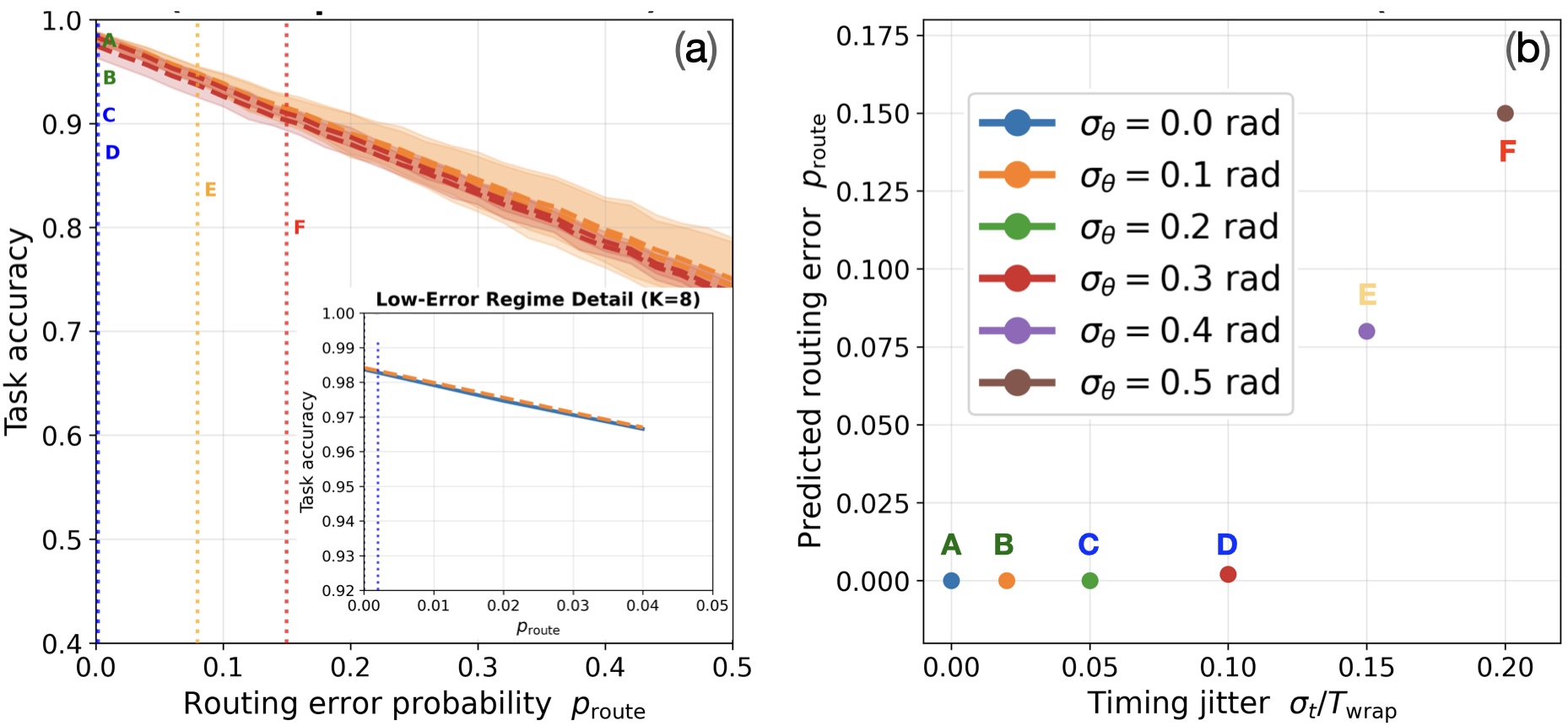}
\caption{
\textbf{MoE routing benchmark linking device-level noise to task-level utility.}
\textbf{(a)} Task accuracy $P_{\mathrm{task}}$ of a top-1 MoE gate versus injected routing-error probability $p_{\mathrm{route}}$.
Each expert is a cluster-specific binary classifier (logistic regression) trained only on its own cluster, so the curves isolate routing errors.
Curves show two expert-set sizes ($K=8$ and $K=32$) and two task difficulties set by the input-cluster spread: $\sigma_x=0.6$ (specialized experts; strong separation) and $\sigma_x=0.9$ (moderate overlap).
Dashed curves indicate mean accuracy over randomized seeds and translucent bands show the corresponding variability across seeds.
Vertical dotted lines labeled A--F mark routing-error levels predicted from representative device operating points.
\textbf{Inset:} zoom of the low-error regime ($p_{\mathrm{route}}\le 0.05$) for $K=8$.
\textbf{(b)} Device-predicted routing error $p_{\mathrm{route}}=1-P_{\mathrm{correct}}$ versus normalized timing jitter $\sigma_t/T_{\mathrm{wrap}}$ for several static phase-disorder levels $\sigma_\theta$ (legend).
The labeled operating points A--F in (b) correspond to the vertical lines in (a), enabling a direct read-off from device noise budgets to task-level accuracy.
}
\label{fig:moe_routing_benchmark}
\end{figure*}

To connect to hardware, we map the address-selection accuracy of the physical module,
$P_{\mathrm{correct}}\equiv\Pr(\hat{k}=k^\ast)$, to the benchmark via
$
p_{\mathrm{route}} = 1-P_{\mathrm{correct}}.
$
Figure~\ref{fig:moe_routing_benchmark} combines these relations: panel~(a) plots $P_{\mathrm{task}}$ versus injected $p_{\mathrm{route}}$
for $K\in\{8,32\}$ and two difficulty settings $\sigma_x\in\{0.6,0.9\}$, while panel~(b) plots the device-predicted $p_{\mathrm{route}}$
as a function of $\sigma_t/T_{\mathrm{wrap}}$ and $\sigma_\theta$ using the WTA simulations of Fig.~\ref{fig:wta_hero}.
The operating points A--F translate directly from device noise budgets to expected task-level accuracy by reading $p_{\mathrm{route}}$ from (b)
and then $P_{\mathrm{task}}$ from (a).
This benchmark is intentionally routing-limited: it does not model soft routing, load balancing, or end-to-end representation learning.

\paragraph{Limitations and extensions.}
The present primitive outputs a single index and does not implement soft weighting (softmax) or learned mixtures.
Extending to top-$k$ selection would require staged selection (e.g., replicated WTA blocks with inhibition/deflation) or nonlinearities supporting controlled multi-winner coexistence.
A full end-to-end training recipe that jointly learns temporal encoders, template libraries, and expert parameters under measured device noise is beyond the scope of this work.

A platform-specific break-even analysis is beyond the scope of this work, but  we  summarize the dominant scaling of the address-selection step for a decision window with $N$ active spikes and $K$ candidate templates (or experts) in Supplemental Sec.~S7 where we  contrast a conventional digitize--compare--select path with polychronous wave scoring followed by WTA digitization.

\section{Outlook: novelty, impact, and open directions}
\label{sec:outlook}

Polychronous wave computing isolates a timing-native hardware primitive: using spike timing as a direct address for discrete selection and routing, rather than as an intermediate analog variable to be reduced later in clocked logic.
This aligns with the LUT-centric shift emphasized in the Spiking Manifesto \cite{Izhikevich2025SpikingManifesto}: the critical operation in timing-based inference is often the timing$\rightarrow$index step (address selection), not dense multiply--accumulate arithmetic.
Within a wrap-free, phase-coherent decision window, relative delays are mapped to phases; a programmable interference stage scores $K$ temporal hypotheses in parallel; and a driven--dissipative winner-take-all stage digitizes the outcome once at readout.
The result is a one-shot, time-domain lookup primitive: a physical argmax that returns a one-hot address from phase-coded spike delays.

This split between linear phase-sensitive scoring and nonlinear competitive selection makes the operating envelope explicit (wrap time, coherence time, and a unified phase-noise budget) and identifies a practical robustness variable: the winner--runner-up margin that predicts boundary-first errors under jitter, static mismatch, or dephasing.
Because margins are directly observable at the outputs, they also provide a natural calibration target: intensity-only hardware-in-the-loop tuning can enlarge post-selection margins to compensate static offsets and drift, while system-level benchmarks translate measured misrouting probability into task-level impact for sparse top-1 routing (e.g., hard MoE gating).

This role differs from common photonic acceleration routes.
Whereas photonic MAC engines accelerate analog linear algebra and many photonic SNN demonstrations emphasize integrate–and–fire dynamics or rate-coded signals, PWC uses waves to compute the discrete decision itself: parallel interference evaluates temporal hypotheses, and competitive dynamics selects a single address, eliminating per-spike timestamping and downstream digital argmax reduction.

The same functional ingredients: a phase reference, programmable coherent superposition, competitive digitization, and reset, can be supplied by several platforms, including exciton--polariton networks and hybrid photonic architectures that combine a programmable linear scorer with a fast nonlinear oscillator/laser WTA bank.
One near-term systems role is a routing or selection coprocessor for LUT-style spiking and sparse routing (e.g., top-1 expert selection), where multiple such modules could be cascaded to form multilayer spike-pattern lookup pipelines.

The main open question is the end-to-end break-even point: whether native-time selection offsets I/O, memory access, calibration cadence, and interconnect costs relative to digitize–compare–select baselines.
Three concrete directions follow:
\begin{itemize}
\item \textbf{Beyond hard top-1:} controlled top-$k$ or multi-winner routing requires nonlinear designs that support reproducible coexistence and a clear accounting of added calibration overhead.
\item \textbf{Joint learning under measured noise:} training temporal encoders, template libraries, and downstream value transforms together using empirically measured margin/noise statistics remains largely unexplored.
\item \textbf{Scaling rules for cascades:} larger compositions will need explicit architectural constraints for managing crowding, preserving coherence, and maintaining margin distributions across stages.
\end{itemize}
A practical way to evaluate time-domain hardware, in this spirit, is by which discrete decisions it can compute before digital abstraction and at what measured routing-error rate; a minimal milestone is a small-$K$ demonstrator that reports routing error versus injected jitter together with margin distributions before/after calibration.

\begin{acknowledgments}
The author acknowledges support from HORIZON EIC-2022-PATHFINDERCHALLENGES-01 HEISINGBERG Project 101114978, from Weizmann--UK Make Connection Grant 142568, and from the EPSRC UK Multidisciplinary Centre for Neuromorphic Computing (grant UKRI982).
\end{acknowledgments}

 \appendix
 \makeatletter
\renewcommand{\thesubsection}{\Alph{section}.\arabic{subsection}}
\renewcommand{\p@subsection}{} 
\makeatother
\section{Additional analytical results}
\label{app:analytics}
\renewcommand{\thesubsection}{\Alph{section}.\arabic{subsection}}

This appendix derives the analytic estimates used in Secs.~\ref{sec:linear_lookup}, \ref{sec:robustness}, and \ref{sec:circuits}.
Supplementary material provides numerical checks (finite-$N$, correlated libraries, and Monte Carlo details).

\subsection{Interferometric score: matched vs.\ random-phase templates}
\label{app:matched_unmatched}

The linear interferometric score (complex field amplitude) for output/template $k$ is
\begin{equation}
\Psi_k
=
\sum_{j=1}^{N} |J_{jk}|\,
\exp\!\left[i\left(\theta_{jk}-\Omega t_j\right)\right],
\label{eq:app_Psi_def}
\end{equation}
with input times $t_j$, angular rate $\Omega$, programmed phases $\theta_{jk}$ (defined modulo $2\pi$), and magnitudes $|J_{jk}|\ge 0$.
We use the amplitude and linear intensity
\begin{equation}
A_k \equiv |\Psi_k|,\qquad
I_k^{(\mathrm{lin})}\equiv |\Psi_k|^2 .
\end{equation}

\paragraph*{Matched template.}
For the matched channel $k^\ast$, compilation sets $\theta_{j k^\ast}=\Omega t_j^{(k^\ast)}$ (mod $2\pi$), ...and a perfect match (no wrap ambiguity) means
$\Omega(t_j-t_j^{(k^\ast)})\equiv 0 \ (\mathrm{mod}\ 2\pi)$ for all $j$.
If spike times are restricted to a single wrap interval (e.g.\ $t_j\in[0,2\pi/\Omega)$), this reduces to $t_j=t_j^{(k^\ast)}$.
 Then
\begin{equation}
\Psi_{k^\ast} = \sum_{j=1}^{N} |J_{j k^\ast}|,
\qquad
I_{k^\ast}^{(\mathrm{lin})}=\left(\sum_{j=1}^{N} |J_{j k^\ast}|\right)^2.
\label{eq:app_matched}
\end{equation}
For uniform magnitudes $|J_{jk}|=1$, this gives $A_{k^\ast}=N$ and $I_{k^\ast}^{(\mathrm{lin})}=N^2$.

\paragraph*{Unmatched template (random-phase baseline).}
As a dephased baseline, assume the residual phases
$\varphi_{jk}\equiv \theta_{jk}-\Omega t_j$ are i.i.d.\ uniform on $[0,2\pi)$ across $j$.
Then $\mathbb{E}[\Psi_k]=0$ and
\begin{equation}
\mathbb{E}[|\Psi_k|^2]=\sum_{j=1}^{N}|J_{jk}|^2,
\qquad k\neq k^\ast,
\label{eq:app_unmatched}
\end{equation}
so the typical (rms) unmatched amplitude scales as
$A_k \sim \sqrt{\sum_j |J_{jk}|^2}$.
For $|J_{jk}|=1$, $A_k\sim \sqrt{N}$ and $I_k^{(\mathrm{lin})}\sim N$.

Here ``typical'' refers to the root-mean-square amplitude:
$\Psi_k=\sum_j |J_{jk}|e^{i\varphi_j}$ with i.i.d.\ $\varphi_j\sim\mathrm{Unif}[0,2\pi)$ gives
\begin{equation}
\mathbb{E}\!\left[|\Psi_k|^2\right]=\sum_j |J_{jk}|^2,
\qquad
|\Psi_k|_{\mathrm{rms}}\equiv \sqrt{\mathbb{E}[|\Psi_k|^2]}.
\label{eq:unmatched_rms}
\end{equation}
The mean amplitude $\mathbb{E}[|\Psi_k|]$ differs only by an $\mathcal{O}(1)$ Rayleigh factor from $|\Psi_k|_{\mathrm{rms}}$ (same scaling in $N$).

\paragraph*{Weight-robust contrast.}
Define the coherent-to-incoherent intensity ratio
\begin{equation}
C_k \equiv
\frac{\left(\sum_j |J_{jk}|\right)^2}{\sum_j |J_{jk}|^2}.
\label{eq:app_contrast_def}
\end{equation}
By Cauchy--Schwarz, $1\le C_k \le N$, with $C_k=N$ iff the magnitudes are uniform.
It is therefore natural to interpret $N_{\rm eff}\equiv C_k$ as an effective coherent count for inhomogeneous weights.

\subsection{Phase noise: jitter, static offsets, and finite coherence}
\label{app:jitter_disorder}

Timing jitter $\delta t_j$ produces rotating-frame phase noise $\delta\phi_j=\Omega\,\delta t_j$.
Static phase offsets act as $\theta_{jk}\mapsto \theta_{jk}+\delta\theta_{jk}$.
Finite coherence over the decision window contributes an additional phase diffusion with variance $\sigma_{\rm coh}^2$ (Sec.~\ref{subsec:jitter_coherence}).
For analytic estimates we treat the matched channel $k^\ast$ as a sum of aligned phasors with additive phase errors
\begin{equation}
\varepsilon_j=\delta\phi_j+\delta\theta_{j k^\ast}+\varepsilon^{(\rm coh)}_j,
\qquad
\mathrm{Var}(\varepsilon_j)=\sigma_{\rm eff}^2,
\end{equation}
with independent zero-mean Gaussians
\begin{equation}
\delta\phi_j \sim \mathcal{N}(0,\sigma_\phi^2),\quad
\delta\theta_{j k^\ast} \sim \mathcal{N}(0,\sigma_\theta^2),\quad
\sigma_\phi=\Omega\sigma_t,
\label{eq:app_phase_noises}
\end{equation}
and
\begin{equation}
\sigma_{\rm eff}^2 \equiv \sigma_\phi^2 + \sigma_\theta^2 + \sigma_{\rm coh}^2 .
\label{eq:app_sigma_eff}
\end{equation}
(For static offsets, expectations should be read as an ensemble average over residual offsets or as a typical suppression when the residual mismatch across $j$ is effectively random.)

\paragraph*{Mean complex amplitude.}
Using $\mathbb{E}[e^{i\varepsilon}]=e^{-\sigma_{\rm eff}^2/2}$ for $\varepsilon\sim\mathcal{N}(0,\sigma_{\rm eff}^2)$,
\begin{equation}
\mathbb{E}[\Psi_{k^\ast}]
=
\left(\sum_{j=1}^{N} |J_{j k^\ast}|\right)\,
e^{-\sigma_{\rm eff}^2/2}.
\label{eq:app_mean_ampl}
\end{equation}

\paragraph*{Mean intensity.}
A direct expansion gives
\begin{align}
\mathbb{E}[I_{k^\ast}^{(\mathrm{lin})}]
&=
e^{-\sigma_{\rm eff}^2}\left(\sum_{j=1}^{N} |J_{j k^\ast}|\right)^2
+\left(1-e^{-\sigma_{\rm eff}^2}\right)\sum_{j=1}^{N} |J_{j k^\ast}|^2.
\label{eq:app_mean_intensity}
\end{align}
For $|J_{jk}|=1$,
\begin{equation}
\mathbb{E}[I_{k^\ast}^{(\mathrm{lin})}]
=
N^2 e^{-\sigma_{\rm eff}^2} + N\left(1-e^{-\sigma_{\rm eff}^2}\right).
\label{eq:app_uniform_intensity}
\end{equation}

\paragraph*{Contrast under phase diffusion.}
In the random-phase baseline, a typical unmatched intensity scale for the same magnitudes is $\sum_j |J_{j k^\ast}|^2$.
The matched-to-unmatched mean intensity ratio is therefore
\begin{equation}
C_{\rm eff}
\equiv
\frac{\mathbb{E}[I_{k^\ast}^{(\mathrm{lin})}]}{\sum_j |J_{j k^\ast}|^2}
=
1 + (C_{k^\ast}-1)e^{-\sigma_{\rm eff}^2}.
\label{eq:app_Ceff}
\end{equation}

\subsection{Extreme-value scaling in the random-phase model}
\label{app:extreme_value}

For $k\neq k^\ast$ in the random-phase baseline (Sec.~\ref{app:matched_unmatched}), let
\begin{equation}
S_k \equiv \sum_{j=1}^{N}|J_{jk}|^2 .
\end{equation}
For large $N$ and no single dominant term, a CLT approximation gives
$\Psi_k \approx X+iY$ with $X,Y\sim\mathcal{N}(0,S_k/2)$, so $|\Psi_k|$ is approximately Rayleigh with tail
\begin{equation}
\Pr(|\Psi_k|>r)\approx \exp\!\left(-\frac{r^2}{S_k}\right).
\label{eq:app_rayleigh_tail}
\end{equation}
For $|J_{jk}|=1$, $S_k=N$ and the tail is $\exp(-r^2/N)$.

Let $M_K \equiv \max_{k\neq k^\ast}|\Psi_k|$ over $K-1$ approximately independent competitors with a common scale $S$.
A union bound gives
\begin{equation}
\Pr(M_K>r)\;\lesssim\; (K-1)\exp\!\left(-\frac{r^2}{S}\right),
\end{equation}
so the typical maximum satisfies
\begin{equation}
M_K \sim \sqrt{S\log K},
\label{eq:app_max_scaling}
\end{equation}
up to order-unity constants (and lower-order $\log\log K$ corrections).

\paragraph*{Typical margin (coherence-dominated matched channel).}
Using \eqref{eq:app_mean_ampl} as the coherent scale for the matched amplitude, a typical linear margin obeys
\begin{eqnarray}
\Delta_{\rm lin}
&=&
A_{k^\ast}-\max_{k\neq k^\ast}A_k \\
&\sim&
\left(\sum_{j=1}^{N}|J_{j k^\ast}|\right)e^{-\sigma_{\rm eff}^2/2}
- c\,\sqrt{S\log K},
\label{eq:app_margin_scaling}
\end{eqnarray}
with $c=\mathcal{O}(1)$ and $S$ the competitor scale (e.g.\ $S=N$ for uniform weights).
The first term is the correct scale when the coherent contribution dominates the incoherent floor, i.e.\
$C_{k^\ast}e^{-\sigma_{\rm eff}^2}\gg 1$ (for uniform weights, $N e^{-\sigma_{\rm eff}^2}\gg 1$).

\subsection{Margin-to-error proxy}
\label{app:margin_error}

Let $\Delta_{\rm lin}=A_{k^\ast}-A_{\rm ru}$ with $A_{\rm ru}\equiv\max_{k\neq k^\ast}A_k$.
As a compact proxy, model the fluctuation of the top-two gap by $\delta\Delta\sim\mathcal{N}(0,\sigma_\Delta^2)$ and define
\begin{eqnarray}
P_{\rm error}&\equiv& \Pr\!\left[\Delta_{\rm lin}+\delta\Delta<0\right]
=
\Phi\!\left(-\frac{\Delta_{\rm lin}}{\sigma_\Delta}\right)\nonumber \\
&=&
\frac{1}{2}\,\mathrm{erfc}\!\left(\frac{\Delta_{\rm lin}}{\sqrt{2}\,\sigma_\Delta}\right),
\label{eq:app_Perror}
\end{eqnarray}
with $\Phi$ the standard normal CDF.
A convenient exponential tail bound for $\Delta_{\rm lin}\ge 0$ is
\begin{equation}
P_{\rm error}\le \frac{1}{2}\exp\!\left(-\frac{\Delta_{\rm lin}^2}{2\sigma_\Delta^2}\right).
\end{equation}
This is used only as a gap-to-noise proxy; reported accuracies come from the full driven--dissipative simulations.

\subsection{Two-spike temporal-order comparator}
\label{app:order_comparator}

For two spikes at times $t_A,t_B$, define rotating-frame phases $\phi_A=\Omega t_A$, $\phi_B=\Omega t_B$, and
$\Delta\phi\equiv \phi_B-\phi_A=\Omega(t_B-t_A)\equiv \Omega\Delta t$.
A minimal two-output interferometric router is
\begin{align}
A_L &= J\!\left(e^{i(\phi_A+\theta_L^{(A)})}+e^{i(\phi_B+\theta_L^{(B)})}\right),\\
A_R &= J\!\left(e^{i(\phi_A+\theta_R^{(A)})}+e^{i(\phi_B+\theta_R^{(B)})}\right),
\end{align}
with $I_{L,R}=|A_{L,R}|^2$ (the global phase of $J$ drops out).

Choose
\begin{equation}
\theta_L^{(A)}=0,\quad \theta_L^{(B)}=-\frac{\pi}{2},\qquad
\theta_R^{(A)}=-\frac{\pi}{2},\quad \theta_R^{(B)}=0,
\end{equation}
which yields
\begin{eqnarray}
I_L &=& 2|J|^2\!\left(1+\sin\Delta\phi\right),\quad
I_R = 2|J|^2\!\left(1-\sin\Delta\phi\right),\nonumber\\
&&I_L-I_R = 4|J|^2\sin(\Omega\Delta t).
\label{eq:II}
\end{eqnarray}
To map sign$(I_L-I_R)$ unambiguously to temporal order, require $|\Delta\phi|<\pi$, i.e.\
\begin{equation}
|t_B-t_A| < \frac{\pi}{\Omega},
\end{equation}
so that $\mathrm{sign}[\sin(\Delta\phi)]=\mathrm{sign}(\Delta t)$.
Supplement Sec.~S4 implements the same $\sin(\Omega\Delta t)$ discrimination in a realizable coupled-mode junction with an equivalent effective loop phase.

\section{Additional candidate platforms and trade-offs}
\label{app:platforms}

\paragraph{Coherent Ising machines (CIMs).}
CIMs realize binary phase states in parametric oscillators and use gain competition plus engineered couplings to select among collective configurations \cite{Marandi2014NatPhoton,Inagaki2016NatPhoton,Inagaki2016Science,McMahon2016Science}.
While CIMs are not designed as spike-latency address selectors, they illustrate scalable competitive dynamics and programmable coupling strategies that could inform multiport WTA implementations.

\paragraph{Magnonic (spin-wave) networks.}
Spin waves provide a coherent carrier with phase as a native variable (spin precession angle), enabling delay-to-phase encoding in a rotating frame.
Programmable interference can be shaped via bias fields, magnonic crystals, or tunable scatterers \cite{Papp2021NatComm,Chumak2015MagnonSpintronics}.
At higher drive powers, nonlinear mode competition and magnon condensation suggest a potential route to WTA digitization \cite{wang2024nanoscale,Demokritov2006MagnonBEC,nowik2012spatially}.

\paragraph{RF and microwave oscillator networks.}
RF/microwave systems naturally provide phase-coherent carriers referenced to a shared clock.
Programmable linear mixing can be implemented with analog phase shifters and reconfigurable networks, including metamaterial processors \cite{Estakhri2019Science}.
Competitive selection can be realized via injection locking or shared-gain saturation; oscillator-network $k$-WTA models have been analyzed in this setting \cite{WangSlotine2004KWTA}.

\paragraph{Acoustic and phononic systems.}
Acoustic/elastic waves support coherent phase evolution and interference in engineered media.
Wave dynamics can be trained for temporal classification during scattering \cite{Hughes2019WaveRNN}, and metamaterials can implement compact phase accumulation and delay elements \cite{zhu2016implementation}.
Digitization would require an added nonlinear or active-gain mechanism (e.g., mode competition in optomechanical ``phonon lasers'') \cite{Vahala2009PhononLaser}.

\bibliography{refsSNN}

\end{document}